\documentclass[lettersize,journal]{IEEEtran}
\usepackage{amsmath,amsfonts}
\usepackage{algpseudocode}
\usepackage{algorithm}
\usepackage{array}
\usepackage{caption}
\usepackage{subcaption}
\usepackage{xcolor}
\usepackage{textcomp}
\usepackage{stfloats}
\usepackage{url}
\usepackage{verbatim}
\usepackage{graphicx}
\usepackage{cite}
\usepackage{booktabs}   
\usepackage{authblk}
\usepackage[switch]{lineno}
\usepackage{censor}

\newcommand{\sect}[1]{Sec.\ref{sec:#1}}

\begin{document}

\title{MMStencil: Optimizing High-order Stencils on Multicore CPU using Matrix Unit}


\author{Yinuo Wang$^*$, Tianqi Mao$^*$, Lin Gan, Wubing Wan, Zeyu Song, Jiayu Fu, Lanke He,\\ Wenqiang Wang, Zekun Yin, Wei Xue, and Guangwen Yang

\thanks{Yinuo Wang$^*$, Tianqi Mao$^*$, Lin Gan, Wubing Wan, Zeyu Song,, Jiayu Fu, Lanke He, Wei Xue, and Guangwen Yang are with the Tsinghua University, Beijing, China}
\thanks{Zekun Yin are with School of Software, Shandong University, Jinan 250100, China}
\thanks{Wenqiang Wang are with High Performance Computing Department, National Supercomputing Center in Shenzhen Shenzhen, China }

}



\maketitle

\def\thefootnote{*}\footnotetext{These authors contributed equally to this work}\def\thefootnote{\arabic{footnote}}

\begin{abstract}
Matrix-accelerated stencil computation is a hot research topic, yet its application to 3 dimensional (3D) high-order stencils and HPC remains underexplored. With the emergence of matrix unit on multicore CPU, we analyze matrix-based accelerating strategies and tailor an optimal approach for 3D high-order stencils. We introduce algorithmic optimizations based on SIMD and matrix unit to address strided memory accesses, alignment conflicts, and redundant accesses. We propose memory optimizations to boost on-package memory efficiency, and a novel multi-thread parallelism paradigm to overcome data-sharing challenges caused by the absense of shared data caches. MMStencil sustains consistently high hardware utilization across diverse stencil shapes and dimensions. Our DMA-based inter-NUMA communication further mitigates NUMA effects and MPI limitations in hybrid parallelism. Combining all the innovations, MMStencil outperforms state‐of‐the‐art libraries on Nividia A100 GPGPU by up to 2.1× . Moreover, the performance improvements enabled by our optimizations translate directly to real‐world HPC applications and enable RTM real-world applications to yield 1.8x speedup versus highly-optimized industrial Nvidia A100 GPGPU version.
\end{abstract}


\section{Introduction}
\begin{figure}[h]
     \centering
     \includegraphics[width=\linewidth]{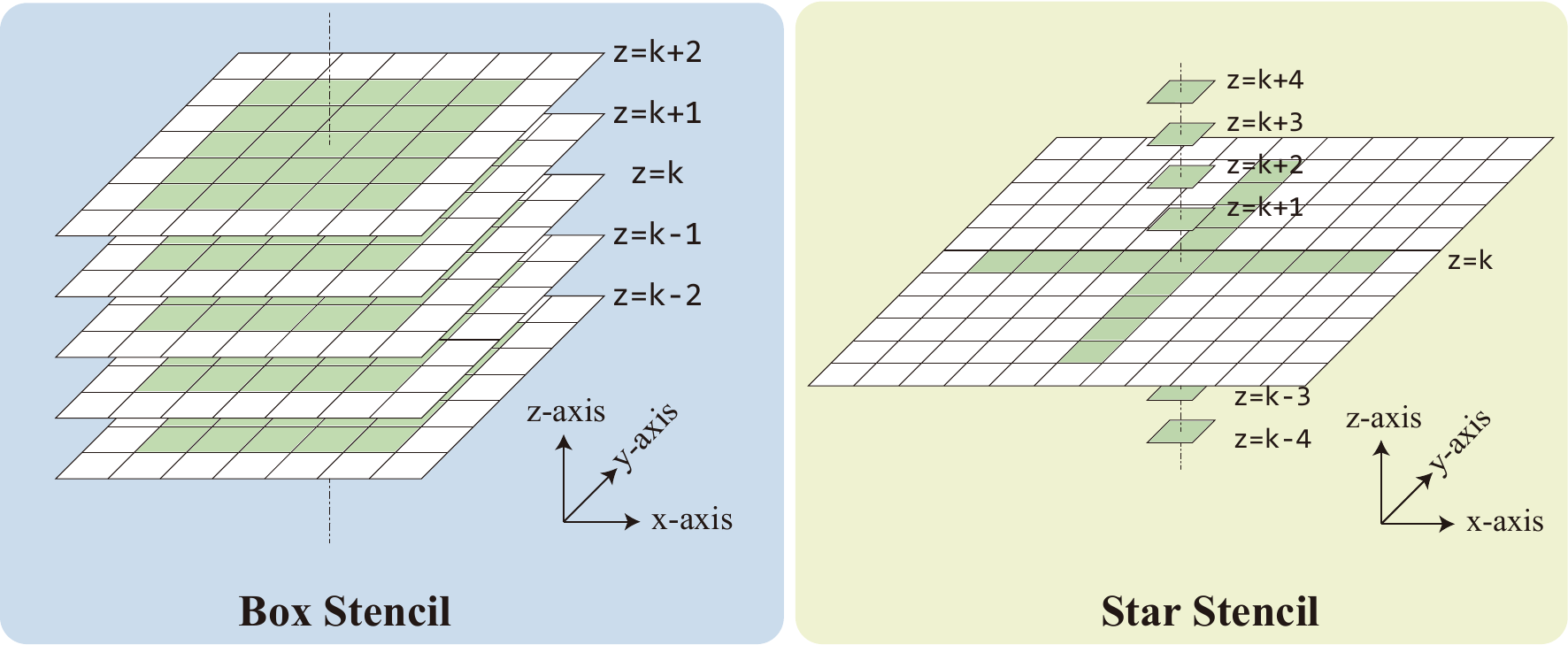}
     \caption{Two Stencil Examples: Box and Star}
     \label{fig:stencil_pattern}
\end{figure}

\IEEEPARstart{S}{tencil} computations are popular study interests in the high performance computing (HPC) community. A stencil computation operates on an N-dimensional grid, updating grid points based on predefined patterns of neighboring cells. 
Fig.~\ref{fig:stencil_pattern} illustrates two widely used stencil patterns: the star stencil, which accesses neighboring cells only along the coordinate axes, and the box stencil, which encompasses all adjacent cells.
Recognized as one of the 13 fundamental HPC \textit{dwarfs}\cite{Asanovic2006, Asanovic2009}, stencil computations underpin computational patterns in finite volume and finite difference methods for discretizing partial differential equations (PDEs), and constitute performance hotspots in applications such as weather forecasting\cite{Chen2014, Ullrich2010}, fluid dynamics\cite{Huynh2013, Lusher2021}, and earth modeling\cite{Chen2018, Zhang2014, Wan2023, Zhou2018}.

Stencil computations manifest in a diverse array of patterns to satisfy differing accuracy and performance requirements—from compact, low-order kernels to expansive, high-order operators. Among these patterns, three-dimensional(3D) high-order stencils form a particularly crucial subclass, underpinning many real-world high-performance computing applications\cite{Ullrich2010, Zhang2014, Chen2018, Wan2023}. For instance, Reverse Time Migration (RTM), a pivotal supercomputing application for earth modeling, employs finite difference methods for approximating spatial derivatives. To balance computational efficiency and storage demands, geophysics researchers frequently utilize high-order stencils within RTM workflows\cite{Zhou2018, Li2017}. A typical scenario involves employing a radius-4 stencil, achieving up to 8-order spatial accuracy. 

From an optimization perspective, stencil computations are typically memory‐bound kernels; consequently, most prior work has focused on exploiting spatial and temporal data locality through blocking strategies to enhance data reuse and reduce redundant memory accesses. 3D high‐order stencils, however, introduce additional challenges: their increased computational workload demands more effective utilization of the processor’s arithmetic units, and their extended stencil radius exacerbates data‐reuse difficulties, calling for more sophisticated techniques to maintain high memory‐bandwidth efficiency.

Given the critical importance of stencil patterns in HPC applications, optimizing the stencil kernel has become a prominent research focus within the community. Moreover, each new hardware architecture inspires renewed efforts to harness its features for ever-more efficient stencil implementations: from multicore CPUs equipped with SIMD instructions\cite{Henretty2011, Henretty2013, Li2021, Li2022, Basu2015, Rawat2018, Yount2016, Alappat2021, Kukreja2016} to GPUs leveraging SIMT parallelism and on-package memory\cite{Nguyen2010, Liu2023, Verweij2017, Vizitiu2014, Falch2014, Rawat2019, Zhao2018, Zhao2019, Zhao2021}. Driven by the rise of AI, GPUs and other high-performance processors now incorporate matrix-acceleration units—most notably tensor cores—that deliver matrix throughput far beyond that of traditional vector units, enabling highly efficient execution of matrix-multiplication kernels at the heart of AI workloads. The remarkable compute density of these units has, in turn, inspired researchers to recast stencil computations as matrix-compatible operations, so that stencil kernels can leverage tensor cores (and similar engines) for substantial acceleration. TCStencil \cite{Liu2022} pioneered the transformation of stencil computations into matrix multiplications, leveraging half-precision tensor cores to accelerate both 2D box and star stencils. ConvStencil \cite{Chen2024} takes an alternative route by first expressing stencils as convolutions and then applying a modified Im2Col operation to improve tensor-core utilization; however, the additional Im2Col overhead limits its overall performance. Building on TCStencil, LoRAStencil \cite{Zhang2024} introduces a low-rank decomposition technique to eliminate redundant shared-memory accesses in 2D box stencils, further boosting efficiency.

Despite their initial promise, subsequent reproduction study\cite{Zhang2025} has revealed that tensor-core–based stencil accelerations seldom outperform well-tuned CUDA-core implementations, tempering the community’s expectations. Furthermore, none of these approaches address three-dimensional high-order stencils. In fact, both CPU-based kernels and GPU-based stencils suffer from low hardware utilization and degraded performance when applied to 3D high-order patterns. 

Another limitation of prior stencil‐optimization efforts is their exclusive focus on the kernel itself, without demonstrating end‐to‐end speedups in real‐world HPC codes. Real‐world HPC workflows often necessitate coupling multiple variables, requiring combinations of different stencil patterns. Additionally, these workflows often involve performing stencil computations on the results of previous stencil operations. In practice, the complexity of real-world applications introduce additional integration challenges that must be overcome to realize true application‐level benefits.

To overcome these limitations, the latest generation of a certain RISC multicore CPU has emerged as a promising alternative. The introduction of the outer-product based matrix unit, combined with integrated on-package memory provide computation power and memory bandwidth comparable with GPGPU. The multi-level cache hierarchy provides opportunities for efficient data reuse. Furthermore, the new low-latency, high-throughput outer-product based matrix computation mechanism offers new opportunities for optimizations, and the programmability inherent to CPUs  simplifies integration and optimization of HPC kernels. These advancements position this RISC multicore CPU as highly competitive alternatives capable of matching or exceeding GPU performance in complex, high-order 3D stencil computations.

In our work, we present MMStencil, a novel matrix-based stencil solution targets at optimizing 3D high-order stencils and real-world applications by harnessing the computational capabilities of the matrix unit alongside system-wide architectural optimizations. Major contributions are as follows:
\begin{enumerate}
    \item We analyze approaches to accelerate stencil computations on matrix acceleration hardware and develop tailored algorithms for matrix unit. Innovative techniques include (1) Tile-Based ILP for Outer-Product Computation, (2) Tile-Assisted Vector Transpose, (3) Cache Pollution Avoiding Intermediate Result Placement, and (4) Redundant-Access Zeroing Box Stencil. With these optimizations, MMStencil sustains high hardware utilization across stencil patterns of varying shapes and dimensions.

    \item We introduce parallel optimizations that exploit the characteristics of RISC multicore SoC to overcome data-sharing challenges caused by the absense of shared data cache, along with memory optimizations for on-package memory to enhance bandwidth utilization. Moreover, we leverage DMA-based communication techniques to mitigate NUMA effects and the limitations of MPI in hybrid parallelism, and employ a pipeline overlapping scheme to achieve higher scalability.
    \item We propose detailed procedures for integrating our stencil kernels into real-world Reverse Time Migration applications, achieving a 2-fold speedup over highly optimized SIMD implementations. Combined with parallel optimizations we achieves up to 3.5-fold speedup over GPU implementations.
\end{enumerate}

\section{Background}
\subsection{High-order Stencil}
High-order stencils arise from the need to estimate spatial derivatives when solving numerical PDEs. In the finite volume (FVM) and finite difference (FDM) methods, derivatives are approximated using neighboring grid points, and increasing the number of points typically enhances accuracy. Consequently, HPC applications that require high precision\cite{Ullrich2010, Zhang2014, Chen2018, Wan2023} often employ high-order stencils with significantly larger radii. Moreover, in scenarios with substantial storage demands\cite{Zhou2018, Li2017, Jacquelin2022}, high-order stencils are frequently used to trade computational intensity for storage constraints.

In wave propagation problems, for an example, accurately modeling a wave with frequency $f_0$ inside a region $(\text{Dim}_X, \text{Dim}_Y, \text{Dim}_Z)$ requires that the number of grid points per wavelength exceed a threshold $G_{\text{stencil}}$ to minimize numerical dispersion\cite{Jiang2019, Xiao1997, Jameson2000}. If the wave velocity is $v$, the number of wavelengths along the x-axis then becomes $n_x = \text{Dim}_x \times\frac{f_0}{v}$, and the required number of sample points in x-direction becomes $n_x G_{\text{stencil}}$. The total grid point count thus becomes $n_x \times n_y \times n_z \times G_{\text{stencil}}^3$. Utilizing high-order stencils significantly reduces the required $G_{\text{stencil}}$ and consequently reduces problem size. Practically, a stencil with radius 1 may require more than 10 points per wavelength, whereas a radius-4 stencil requires no more than 5, achieving approximately an eightfold reduction in total grid points. Therefore adopting high-order stencils is an established industry practice in storage-critical HPC applications like RTM.

Furthermore, complex HPC applications often involve coupling multiple variables, necessitating combinations and sequences of different stencils. Again in example of RTM for Vertical Transverse Isotropic (VTI) media\cite{Yunyue2012,Vladimir2016} commonly encountered in sedimentary formations, the governing equations are:
\begin{align*}
    \frac{\partial^2 \sigma_H}{\partial t^2} &= V_p^2 \left\{ (1 + 2\epsilon)\left[ \frac{\partial^2 \sigma_H}{\partial x^2} + \frac{\partial^2 \sigma_H}{\partial y^2} \right] + \sqrt{1 + 2 \delta}\frac{\partial^2 \sigma_V}{\partial z^2} \right\} \\
    \frac{\partial^2 \sigma_V}{\partial t^2} &= V_p^2 \left\{ \sqrt{1 + 2 \delta}\left[ \frac{\partial^2 \sigma_V}{\partial x^2} + \frac{\partial^2 \sigma_V}{\partial y^2} \right] + (1 + 2\epsilon)\frac{\partial^2 \sigma_H}{\partial z^2} \right\}
\end{align*}
where $\sigma_H$ and $\sigma_V$ represent horizontal and vertical stress components, respectively. These equations involve stencil computations on system variables coupled with spatially varying medium properties.

In even more complex cases, such as Tilted Transverse Isotropic (TTI) media\cite{Li2017} frequently observed in thrust-fold belts, the governing equations become increasingly intricate:
\begin{align*}
    \frac{\partial^2 p}{\partial t^2} &= v_{px}^2 H_2 p + \alpha v_{pz}^2 H_1 q + v_{sz}^2 H_1 (p - \alpha q) \\
    \frac{\partial^2 q}{\partial t^2} &= \frac{v_{pn}^2}{\alpha} H_2 p + v_{pz}^2 H_1 q - v_{sz}^2 H_2\left(\frac{1}{\alpha} p - q\right)
\end{align*}
with operators $H_1$ and $H_2$ involving all six second-order partial derivatives:
\begin{align*}
    H_1 &= \sin^2\theta \cos^2 \phi \, \frac{\partial^2}{\partial x^2} +
           \sin^2\theta \sin^2 \phi \, \frac{\partial^2}{\partial y^2} +
           \cos^2\theta \, \frac{\partial^2}{\partial z^2} \\
        &\quad + \sin^2\theta \sin 2\phi \, \frac{\partial^2}{\partial x \partial y} +
           \sin 2\theta \sin \phi \, \frac{\partial^2}{\partial y \partial z} \\
        &\quad +
           \sin 2\theta \cos \phi \, \frac{\partial^2}{\partial x \partial z} \\
    H_2 &= \frac{\partial^2}{\partial x^2} + \frac{\partial^2}{\partial y^2} + \frac{\partial^2}{\partial z^2} - H_1
\end{align*}

Computing these complex kernels requires stencil operations on both primary system variables and intermediate results from preceding stencils, followed by scalar operations to produce final results. The challenge therefore lies not only in handling high-order stencils efficiently but also in integrating optimized kernels seamlessly into complex HPC workflows.

\subsection{Matrix Unit and Multicore SoC} \label{sec:arm9a}
\begin{figure}[h]
     \centering
     \includegraphics[width=\linewidth]{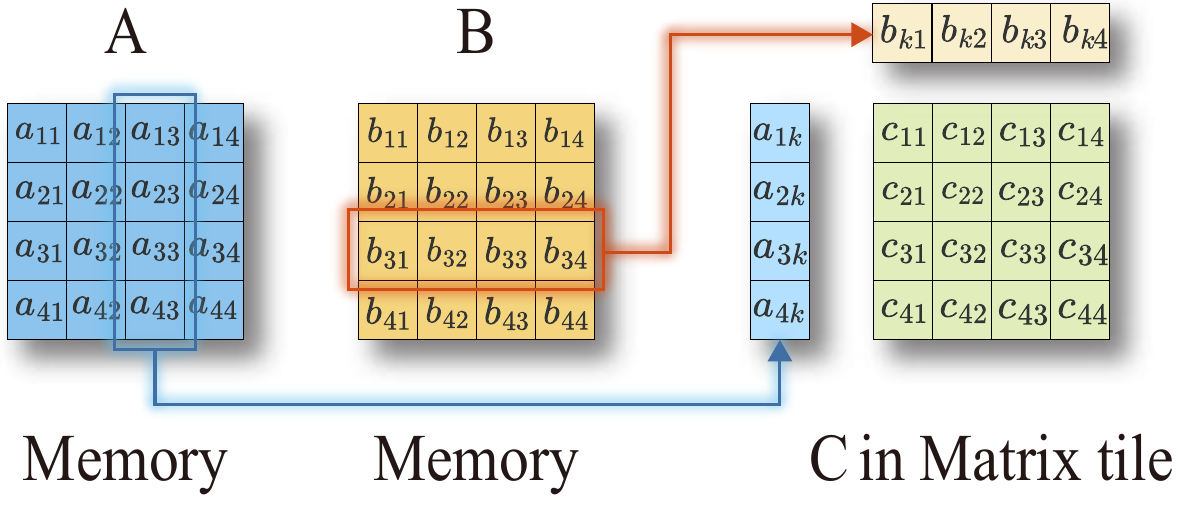}
     \caption{Outer-Product Based Matrix Compute Mechanism.}
     \label{fig:sme_out_product}
\end{figure}

RISC multicore CPUs are renowned for their energy efficiency and multicore parallelization, making them the architecture of choice in world-leading supercomputers such as Fugaku and Tianhe-3. The latest generation of RISC CPUs specifically targets the AI market, and introduces the Matrix unit to accelerate matrix multiplication, which is a key computational hotspot in AI workloads. Unlike the small-fragment matrix units of TPUs or GPUs, the Matrix unit adopts a novel vector outer-product approach. As illustrated in Fig.~\ref{fig:sme_out_product}, each iteration loads a vertical strip of matrix A and a horizontal strip of matrix B into SIMD vector registers, computes outer-product operations, and accumulates the results in a specialized matrix accumulator. This design simplifies programming and facilitates overlap between matrix computations and other tasks on multicore CPUs. Additionally, the Matrix unit provides full floating-point precision support, offering a notable advantage over GPU tensor cores. Many HPC applications, such as RTM, rely heavily on single precision, making the Matrix unit particularly appealing.

The matrix accumulator is an architectural register state consisting of a two-dimensional tile of [64 × 64] bytes in 512-bit SIMD architecture.
For single-precision floats, the tile is partitioned into four independent matrix tiles of size $16\times16$, enabling flexible extraction and insertion of vertical or horizontal slices from/to SIMD vector registers. High-performance matrix multiplication on Matrix unit hinges on interleaving computations across these matrix tiles to hide outer-product latency.

The continuous scaling of core count in multicore CPUs has introduced unique and unprecedented features at the System-on-Chip (SoC) level, opening new opportunities for parallel optimizations. Our chosen experimental CPU platform\footnote{This platform is currently under confidentiality restrictions, so detailed hardware specifications cannot be disclosed. Nevertheless, our experimental results (presented in \sect{experiments}) include both absolute values and percentages of peak performance to ensure interpretability.} 
is a multi-die SoC integrating two compute dies. Each CPU core is equipped with SIMD vector unit, Matrix unit, and private data and instruction caches. Within a single NUMA domain, more than 32 cores are connected in a ring topology and share a finite-capacity on-package. This on-package memory system can be configured in cache mode—serving as a last-level data cache—or in flat mode—functioning as an independent addressable memory region. One compute die integrates four on-package memory NUMA nodes, collectively sharing an off-die DDR memory subsystem that supports up to 1 TB capacity and 120 GB/s bandwidth. A server node can include up to two CPUs. Data traffic between the on-package memory and DDR memory can be explicitly managed by an SDMA engine on each compute die, providing 160 channels for efficient DDR/on-package memory data transfers.

\section{Related Work and Motivation}

\begin{figure*}[h]
     \centering
     \includegraphics[width=\linewidth]{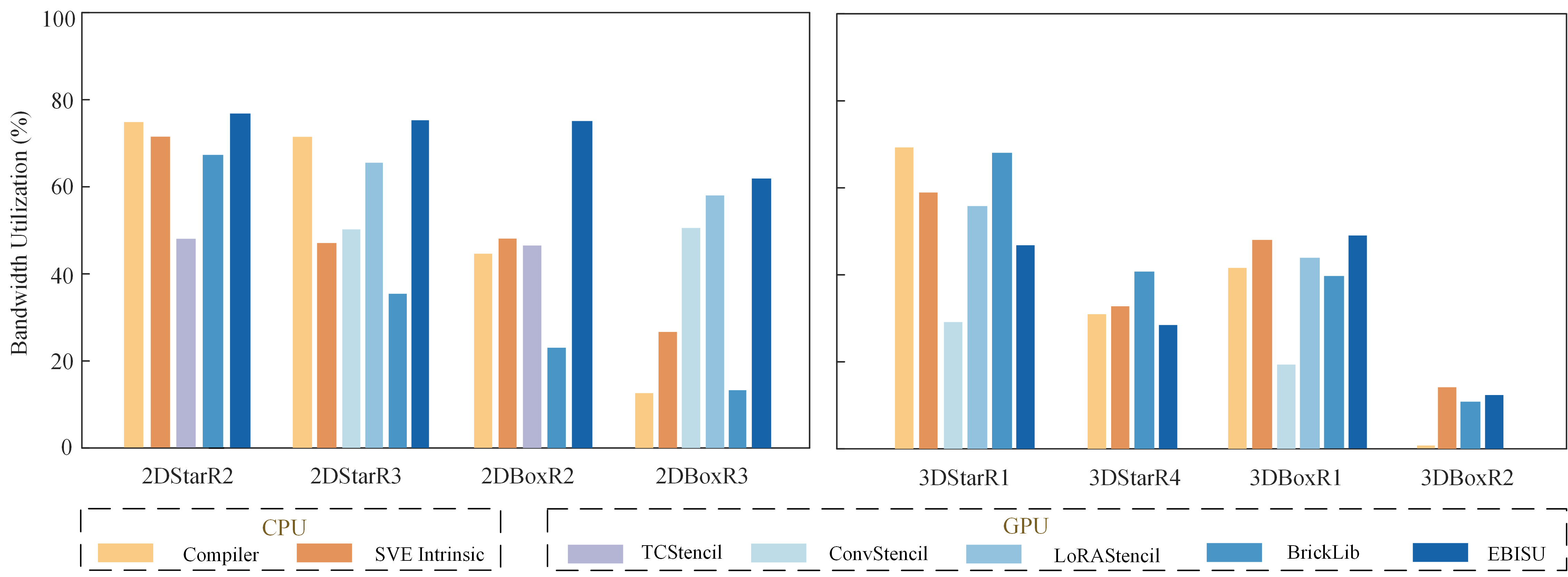}
     \caption{Bandwidth Utilization of State-of-the-arts}
     \label{fig:util_sota}
\end{figure*}

\subsection{Related Work}
Stencil computation is a critical computational pattern that has received considerable attention from the HPC community \cite{Asanovic2009}. Its performance is strongly influenced by factors such as space tiling, vectorization, and hardware characteristics including cache policies and hardware prefetch mechanisms \cite{Datta2009, Nguyen2010}.

\paragraph{CPU-based Stencil Optimization}
Vectorization is essential for achieving high performance. Some approaches focus on exploiting reuse via common sub-expression elimination or enhancing data reuse at the register or cache level \cite{Basu2015, Rawat2018}, while others address data alignment conflicts through memory reordering techniques, such as DLT or vector transpose \cite{Henretty2011, Henretty2013, Li2021, Li2022}. BrickLib leverages data reuse within small blocks and applies memory reordering operations to achieve high-performance stencil computations on both CPUs and GPUs \cite{Zhao2018, Zhao2019, Zhao2021}, inspiring our choice of a brick layout to improve bandwidth utilization. Furthermore, frameworks such as Yask \cite{Yount2016, Alappat2021} and Devito \cite{Kukreja2016} extend beyond kernel-level optimization to accelerate full-scale HPC applications.

\paragraph{GPU-based Stencil Optimization}
On GPUs, tiling techniques are widely used to improve performance \cite{Nguyen2010, Zhao2018, Liu2023, Verweij2017, Vizitiu2014}. Given the abundant register resources on GPUs, many works also exploit register-level data reuse \cite{Falch2014, Rawat2019}. Due to the complexity of developing efficient CUDA kernels, several studies have turned to domain-specific languages (DSLs) to automatically generate highly optimized stencil code. Notable examples include Physis \cite{Maruyama2011}, Lift \cite{Hagedorn2018}, Artemis \cite{Rawat2019}, and AN5D \cite{Matsumura2020}. Although temporal fusion is frequently applied to further boost performance, this approach can reduce the adaptability of the resulting code for HPC applications.

\paragraph{Tensor Core-based Stencil Optimization}
Recent efforts have exploited the computational power of tensor cores to develop stencil kernels. TCStencil \cite{Liu2022} transforms stencil computation into matrix multiplication using a finite-difference approach, while ConvStencil \cite{Chen2024} employs a revised Im2Col transformation to improve tensor core utilization. TCStencil's successor LoRAStencil \cite{Zhang2024} exploits symmetry in box stencils; it uses rank decomposition to transform a box stencil into a series of 1D stencils. This technique trades minor additional computation for maximal data reuse at the shared-memory level and currently represents the state-of-the-art. Nonetheless, these approaches do not address the optimization of 3D high-order stencils or the integration of stencil kernels into  HPC applications.

\subsection{Motivation Experiments}
In this section we evaluate the performance of several SOTA stencil libraries on GPU and our multicore CPU, including the recently very hot tensor-core-based libraries. We basically follow the experimental setting in \cite{Zhang2025}. We aim at evaluating their bandwidth utilization and performance on 3D high-order stencils.

\paragraph{\textbf{Machine}} Our GPU platform is an NVIDIA A100 GPGPU equipped with 80 GB of on-package memory, capable of delivering up to 1955 GB/s of memory bandwidth. Details of the CPU platform are provided in \sect{experimental_setup}.

\paragraph{\textbf{State-of-the-Art}} For GPUs, we evaluate three tensor-core–based libraries (TCStencil \cite{Liu2022}, LoRAStencil \cite{Zhang2024}, and ConvStencil \cite{Chen2024}) alongside two CUDA-core–based libraries (BrickLib \cite{Zhao2018,Zhao2019,Zhao2021} and EBISU \cite{Zhang2023}). We use the same baseline setting provided in \sect{experimental_setup}. To mirror real-world HPC workflows, in which boundary-condition handling often constrains the depth of temporal blocking, we restrict temporal blocking to a single timestep in all implementations.  

\paragraph{\textbf{Benchmarks}} We use the same benchmark settings as in \sect{experimental_setup}. Since most GPU libraries provide only a 3DStarR1 implementation and lack 3DStarR2, we evaluate 3DStarR1 in place of 3DStarR2.

\paragraph{\textbf{Metric}} All GPU libraries execute in double precision, except TCStencil which operates in half precision, while CPU libraries run in single precision. To compare hardware utilization uniformly, we use bandwidth utilization as the metric, defined as
\[
\text{Bandwidth Utilization}
= \frac{2 \times \mathrm{sizeof}(\text{datatype}) \times \text{GStencils}}
       {\text{PeakBandwidth}}.
\]

Figure~\ref{fig:util_sota} presents the hardware utilization of various stencil libraries on both CPU and GPU platforms. The results confirm that prior tensor‐core–based approaches failed to improve utilization. Among the CUDA‐based libraries, EBISU and BrickLib achieve superior bandwidth efficiency across diverse stencil shapes and dimensions. On the CPU side, both the compiler‐optimized baseline and the hand‐tuned SIMD implementation already deliver very high utilization for 2D star stencils of all radii and for 3D stencils with short radii. Therefore on stencil patterns where SIMD and compiler approaches already excel, our work focuses on delivering consistently high performance.

However, for high‐order stencils, all existing libraries suffer severe performance degradation. On the CPU platform, the compiler‐optimized baseline slows by a factor of 2.25× when the 3D star stencil radius increases from 1 to 4, and the hand‐tuned SIMD version by 1.80×. On the GPU platform, BrickLib’s utilization drops by 1.70× and EBISU’s by 1.65×. For box stencils, the reduction in bandwidth efficiency is even more pronounced. We aim to close this performance gap for 3D high‐order stencils by leveraging the Matrix unit’s high computational throughput.

\section{Design of MMStencil}
In this section we present the key aspects of our design and optimizations of our MMStencil framework, including choice of algorithm to map stencil to Matrix unit, micro-architectural level optimization, multi-thread level optimization, multi-process level optimization, and memory optimization. We will also show how to integrate our highly optimized kernel into complex HPC kernel using examples.

\subsection{Mapping Stencil to the Matrix Unit}
The Matrix unit of our multicore CPU accelerate matrix multiplication in a outer-product approach. Given column vector $\boldsymbol{\alpha} = (\alpha_0, \alpha_1, \cdots \alpha_{n-1})^T$ and row vector $\boldsymbol{\beta} = (\beta_0, \beta_1, \cdots \beta_{n-1})$, the outer-product between $\boldsymbol{\alpha}$ and $\boldsymbol{\beta}$ can be interpreted in two equivalent ways: (1) scaling $\boldsymbol{\beta}$ with $\alpha_i$ and accumulate to i'th row, or (2) scaling $\boldsymbol{\alpha}$ with $\beta_j$ and accumulate to j'th column. 

\begin{figure}[h]
     \centering
     \includegraphics[width=\linewidth]{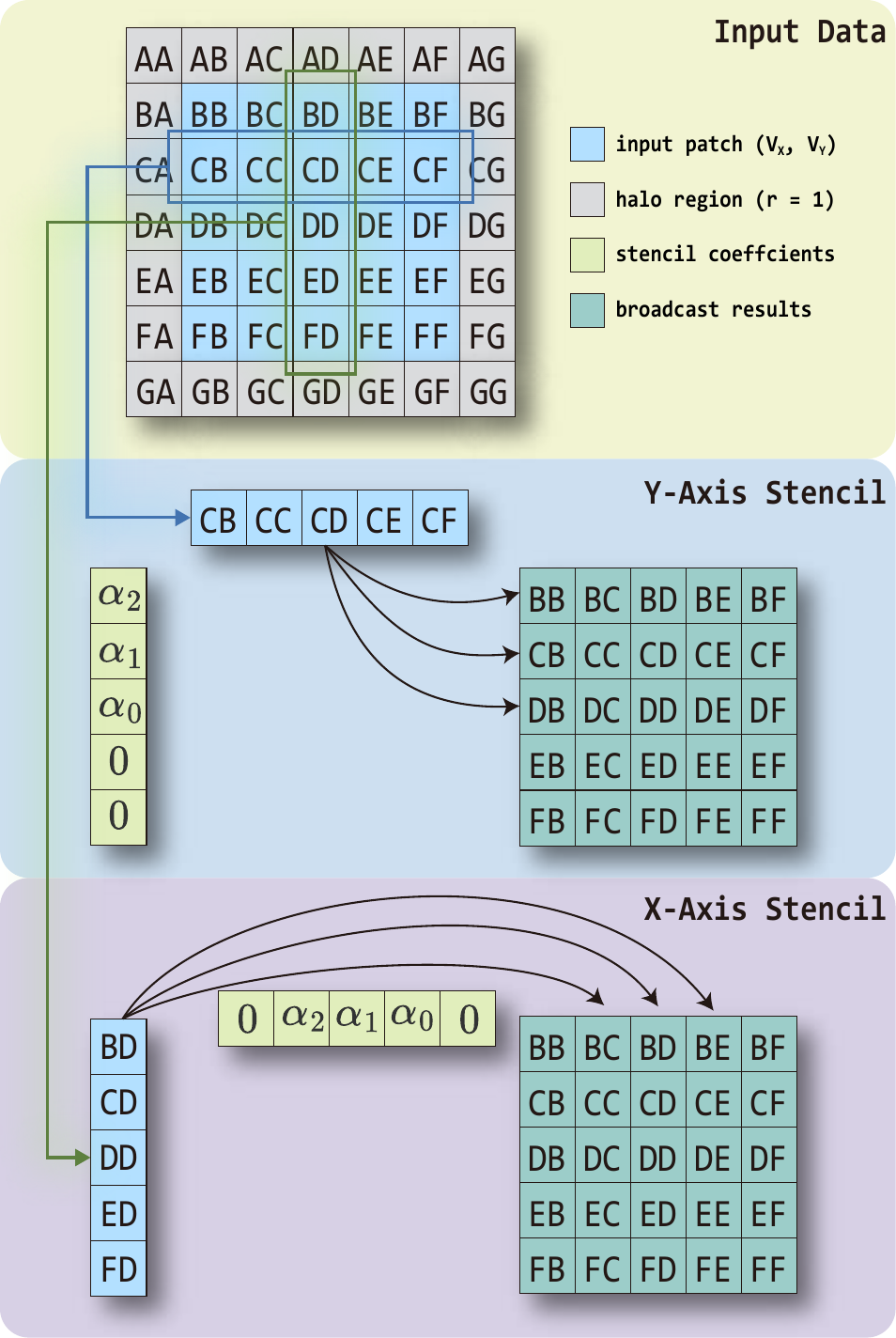}
     \caption{Mapping Stencil to Matrix Unit}
     \label{fig:sme_stencil}
\end{figure}

As shown in Fig.~\ref{fig:sme_stencil}, a one‐dimensional stencil along the y–axis with radius $r$ can be mapped directly to this model. Let the Tile accumulator hold a block of shape $(V_X, V_Y)$; the corresponding input patch must include a halo of width $r$ on each side in the $y$–dimension, i.e.\ an array of size $(V_X, V_Y + 2r)$. We load each $(V_X,  1)$ row vector from the input and broadcast it across all $V_Y$ output rows, weighted by the appropriate stencil coefficients. This is equivalent to performing an outer product between the loaded row and a coefficient‐filled column vector (with zeros in non‐dependent positions). 

For x-axis stencil, the input grid is of shape $(V_X + 2r, V_Y)$ and each $(1, V_Y)$ column vector is scattered to all output columns with appropriate stencil coefficients. The same pattern applies to z–axis stencils by treating the outermost dimension analogously. Complex three‐dimensional stencils can then be realized by composing these 1D outer‐product mappings for the $x$–, $y$–, and $z$–axes in sequence, fully leveraging Matrix unit’s high‐throughput pipelines.

\subsection{A Preliminary Performance Model} \label{sec:perf_model}

In a processor with SIMD vector length $V_L$ and a one-dimensional stencil of radius $r$, computing an output block of shape $(V_L, V_L)$ requires $V_L \times (2r+1)$ SIMD operations. By contrast, the Matrix unit performs $V_L + 2r$ outer-product instructions. Let the cycle‐per‐instruction (CPI) of the SIMD FMA and the Matrix outer-product be $\text{CPI}_{\text{SIMD}}$ and $\text{CPI}_{\text{Matrix}}$, respectively. The total cycles spent on each are
$
\text{Cycles}_{\text{SIMD}} = V_L \times (2r+1) \times \text{CPI}_{\text{SIMD}},
$ and $
\text{Cycles}_{\text{Matrix}} = (V_L + 2r) \times \text{CPI}_{\text{Matrix}}.
$
Therefore, the achievable floating‐point throughput of MMStencil can be expressed as
\[
\mathrm{FLOPS}_{\mathrm{MMStencil}}
\;=\;
\frac{V_L \,(2r + 1)\,\mathrm{CPI}_{\mathrm{SIMD}}}
     {(V_L + 2r)\,\mathrm{CPI}_{\mathrm{Matrix}}}
\;\times\;
\mathrm{FLOPS}_{\mathrm{SIMD}}.
\]

On most modern CPUs, $\text{CPI}_{\text{SIMD}} = 0.5$, and on our experimental platform $\text{CPI}_{\text{Matrix}} = 2$ in single precision.  Therefore, for any stencil radius $r > 1$, Matrix unit delivers a clear performance advantage over the traditional SIMD‐based approach. For example, at $r = 4$, Matrix unit achieves a theoretical 1.5× speedup compared to the SIMD implementation; although this gain may seem modest, our experimental results reveal substantially higher improvements in practice, which we will discuss later.

\subsection{Microarchitectural Optimizations}
In this subsection, we present our microarchitectural optimizations tailored to the multicore CPU architecture. We leverage key features such as Out-of-Order Execution (OOE), the Least Recently Used (LRU) cache policy, SIMD capabilities, and Matrix unit.

\paragraph{\textbf{Tile-Based ILP for Matrix Unit}}
As discussed in Section~\ref{sec:arm9a}, the high throughput of Matrix unit depends on interleaving outer-product computations across multiple matrix tiles. For the $x$- and $y$-axis stencil computations, we process a block of size $(V_X, V_Y, V_Z)$, where $V_X = V_Y = V_L$, and $V_L$ is the vector length; $V_Z$ is a multiple of the number of matrix tiles. Each matrix tile then computes a $(V_X, V_Y, 1)$ slice of the block, and we interleave computations across different layers. As adjacent outer-product operations have no data dependency, the CPU’s OOE hardware can schedule these instructions to achieve high ILP. For the $z$-axis stencil, each tile computes a $(V_X, 1, V_Z)$ slice of the block.

\begin{figure}[h]
     \centering
     \includegraphics[width=\linewidth]{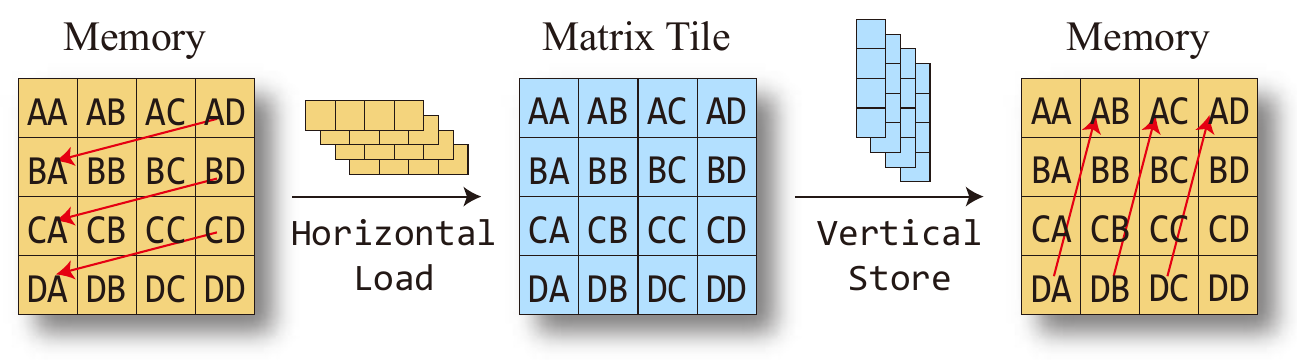}
     \caption{Tile-Assisted Vector Transpose}
     \label{fig:za-transpose}
\end{figure}

\paragraph{\textbf{Tile-Assisted Vector Transpose}}
When computing the $x$-axis stencil, accessing each $(1, V_Y, 1)$ column vector incurs non-contiguous memory accesses. As most modern CPUs perform 2 loads and 1 store per cycle, on a 512-bit SIMD platform, gathering a single-precision vector can require up to 8 cycles. To eliminate this penalty, we apply an explicit transpose to each $(V_X, V_Y, V_Z)$ block along the $xy$-plane. A SIMD permutation-based transpose is costly, requiring $V_L \log_2(V_L)$ permutations. For 512-bit lanes in single precision, this involves up to 64 permutation instructions plus additional load/store instructions. We exploit the ability of matrix tile to extract or insert vertical or horizontal slices from and into the matrix tile, and propse an efficient transpose scheme with a single horizontal load into the matrix tile and one vertical store to memory, as shown in Fig.~\ref{fig:za-transpose}. Only 32 instructions are required for this operation.

\paragraph{\textbf{Cache Pollution Avoiding Intermedian Result Placing}}
In the 3D star stencil, the $x$-axis, $y$-axis, and $z$-axis stencil blocks have incompatible dimensions. For $x$-axis, $y$-axis stencil each matrix tile contains result of shape $(V_X, V_Y, 1)$ and for $z$-axis stencil each matrix tile contains result of shape $(V_X, 1, V_Z)$. Therefore, after computing the $x$- and $y$-axis stencils, partial results are written back to memory and reloaded for the $z$-axis stencil. A subtle but crucial optimization is to use a temporary buffer rather than directly writing to the destination grid. Modern CPUs often implement an LRU cache policy, writing results to the final destination triggers an extra read/write sequence, potentially polluting the cache and impairing performance.

\begin{figure}[h]
     \centering
     \includegraphics[width=\linewidth]{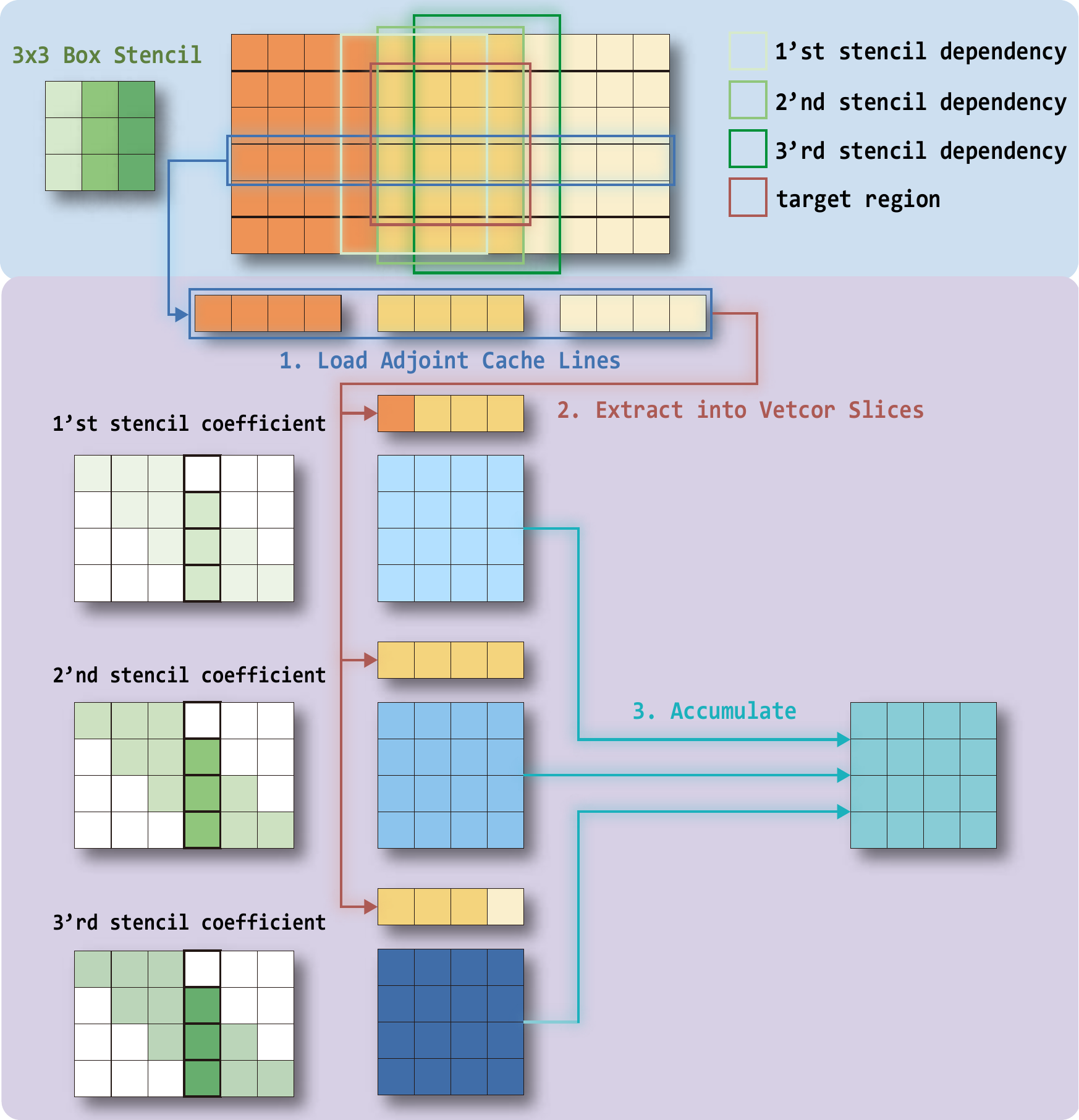}
     \caption{Redundant-Access Zeroing Box Stencil}
     \label{fig:box_stencil_opt}
\end{figure}

\paragraph{\textbf{Redundant-Access Zeroing Box Stencil}}
For a 2D box stencil, the computation is decomposed into $2r+1$ one-dimensional stencils along the $y$-axis. Specifically, the $j$-th stencil must load grid data from \((-j, -r)\) to \((V_X-j, V_Y+r)\), as illustrated in Fig.~\ref{fig:box_stencil_opt}. This decomposition can lead to redundant accesses to the input grid, and when the point \((0, 0)\) is aligned with a cache line, each 1D stencil may suffer from unaligned memory accesses.

In our work, we observe that these $2r+1$ stencils share three adjacent cache lines in each Matrix unit outer product iteration. To exploit this locality, we restructure the computation by moving the iteration over the $y$-axis stencils into the inner loop while executing the Matrix unit stencil procedure in the outer loop, as shown in Fig.~\ref{fig:box_stencil_opt}. SIMD vector splicing is then used to extract the required inputs for each outer product. This approach minimizes redundant memory accesses and improves computational efficiency without incurring extra arithmetic overhead or imposing restrictions on the stencil properties.

\subsection{Memory Optimizations}
Stencil computations demand both high computational power and bandwidth. With our microarchitectural optimizations, the Matrix unit can deliver substantial computational throughput, so our focus shifts to memory optimizations. While on-package memory provides extremely high bandwidth, it widens the data port from 64\,bits (in DDR memory) to 1024\,bits, posing a significant challenge for achieving full utilization. We address this with two approaches: \emph{SIMD-Friendly Memory Reorder} and \emph{Gather-Based Software Prefetch}.

\begin{figure}[h]
     \centering
     \includegraphics[width=\linewidth]{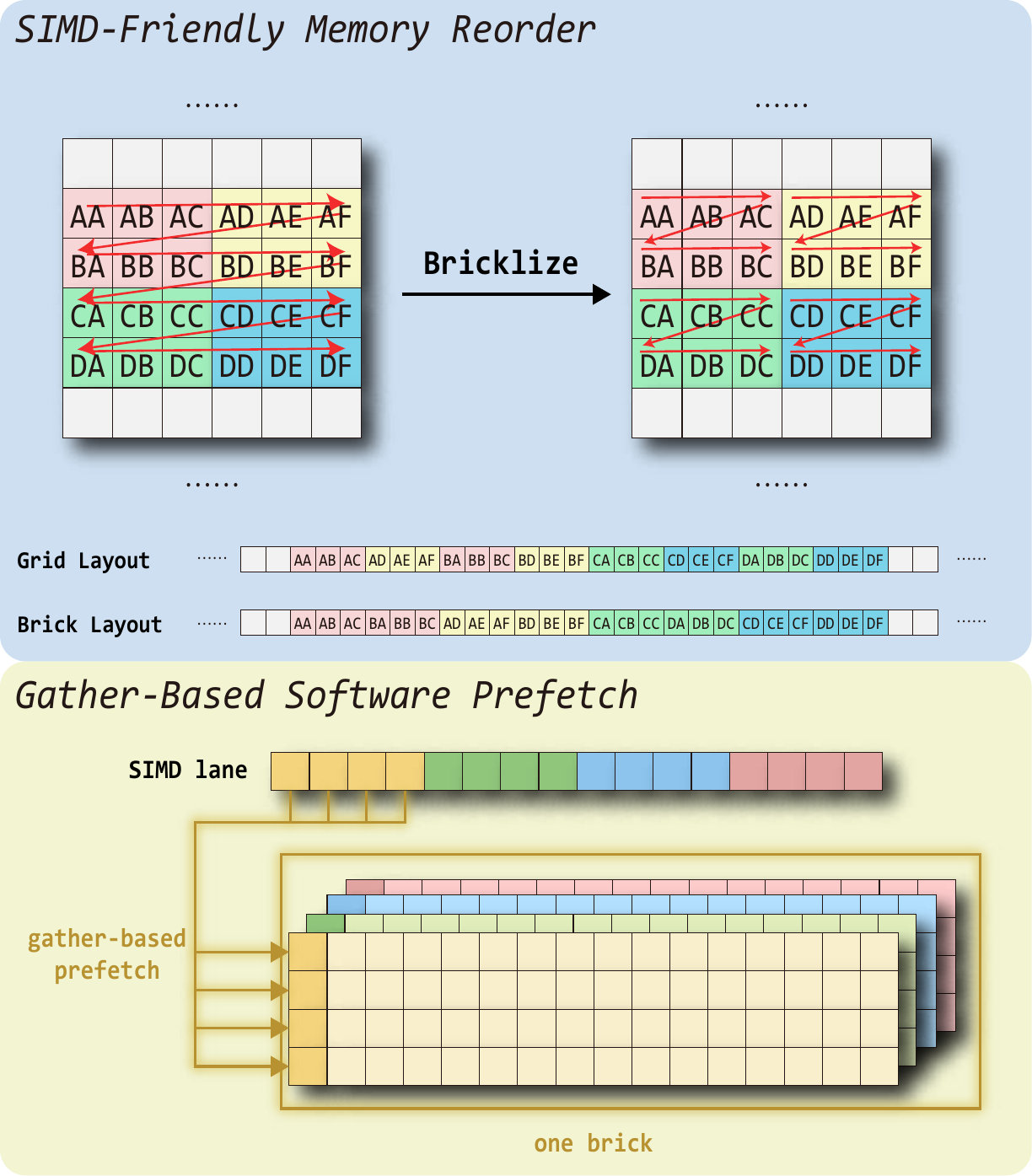}
     \caption{SIMD-Friendly Memory Reorder and Gather-Based Software Prefetch}
     \label{fig:brick_and_prefetch}
\end{figure}

\paragraph{\textbf{SIMD-Friendly Memory Reorder}}
Fully exploiting on-package bandwidth requires carefully managing memory-access streams so that they remain mostly contiguous and few in number. However, under our ``Tile-Based ILP for Matrix Unit'' strategy, we partition the grid into $(V_X, V_Y, V_Z)$ blocks. In 3DStarR4 with single-precision data and $V_X = V_Y = 16$, $V_Z = 4$, this creates $(16 \times 4 \times 3 + 4 \times 4 \times 2) = 226$ distinct memory-access streams, severely degrading on-package memory performance. To mitigate this, we reorder the grid into $(B_X, B_Y, B_Z)$ bricks following schemes from BrickLib\cite{Zhao2018, Zhao2019, Zhao2021}, loading the entire brick whenever the halo region intersects it. Fig.~\ref{fig:brick_and_prefetch} illustrates a 2D example. After reordering the memory layout, the tiled stencil computation exhibits significantly fewer distinct memory‐access streams when processing a $(V_X, V_Y, V_Z)$ block, allowing the kernels to capitalize on the enhanced memory efficiency. For balancing contiguous memory accesses against additional halo traffic, we set $B_X = V_L$ and $B_Y = B_Z = 4$, where $4$ is the largest radius encountered in typical HPC stencils and a divisor of $V_X, V_Y,$ and $V_Z$.

\paragraph{\textbf{Gather-Based Software Prefetch}}
Effective prefetching is critical for overlapping computation with memory accesses. Because our multicore CPU cores lack the sophisticated hardware prefetchers found on x86 CPUs, we introduce software prefetch instructions in MMStencil. Using native vector prefetch for 64\,B requires mixing prefetch instructions within normal stencil computations, which complicates instruction scheduling; if not handled carefully, the overhead can degrade performance. Instead, we exploit the fact that caches operate on a cacheline granularity, using \emph{gather-based prefetch} instructions. Each instruction gathers the head of a cacheline in every slot of the SIMD lane, allowing a single prefetch to fetch $V_L$ cachelines at once. As shown in Fig.~\ref{fig:brick_and_prefetch}, in single precision a single gather‐prefetch instruction suffices to load an entire brick. This approach overlaps computation and memory accesses with minimal instruction overhead.

\subsection{Multi-thread Scope Optimizations}

Multi-core CPUs achieve high performance primarily through multi-core parallelism. With the rapid development of HPC middleware, parallelism can now be efficiently implemented using simple OpenMP directives or parallel libraries such as TBB or SYCL, enabling tasks to be partitioned and dispatched across CPU cores.

However, collaboration and data sharing among cores are often overlooked, with researchers typically delegating this responsibility to hardware mechanisms like the shared Last-Level Cache (LLC). Unfortunately, LLCs exhibit scalability issues as core counts increase. As a result, LLC is absent on platforms such as our multicore CPU platform, with each core only having  limited-capacity private data caches. 

In stencil kernels, where data reuse is critical, the absence of a shared LLC leads to excessive redundant memory traffic and significant performance degradation. In a tiled parallel execution, each core is assigned a tile of shape \((\mathrm{Tile}_X, \mathrm{Tile}_Y, \mathrm{Tile}_Z)\). Halo traffic in each dimension is \(B_X\), \(B_Y\), and \(B_Z\), so the processor must operate on an input block of shape
\(
(\mathrm{Tile}_X + 2B_X,\;\mathrm{Tile}_Y + 2B_Y,\;\mathrm{Tile}_Z + 2B_Z).
\)
Since the outermost (\(z\)) dimension is typically processed iteratively, each \((\mathrm{Tile}_X + 2B_X,\;\mathrm{Tile}_Y + 2B_Y,\;V_Z + 2B_Z)\) slice must fit entirely in a core’s private data cache. The data‐reuse ratio is therefore
\begin{align*}
    &\max_{\mathrm{Tile}_X,\mathrm{Tile}_Y}\quad 
    \frac{\mathrm{Tile}_X \times \mathrm{Tile}_Y}{(\mathrm{Tile}_X + 2B_X)\,(\mathrm{Tile}_Y + 2B_Y)} \\
    &\text{s.t.}\quad
    (V_Z + 2B_Z)\,(\mathrm{Tile}_X + 2B_X)\,(\mathrm{Tile}_Y + 2B_Y)
    \;\le\;
    \mathrm{SIZE}_{\mathrm{LLC}}
    \label{eq:2.5Dblocking}
\end{align*}
where we neglect factors of \(\mathrm{Tile}_Z\) since \(\mathrm{Tile}_Z\) is usually much larger.

Assuming \(\mathrm{Tile}_X\) and \(\mathrm{Tile}_Y\) are powers of two, fitting each tile in a core’s limited private caches typically caps the reuse ratio at around 50\%, leaving nearly one‐third of memory traffic redundant.

\begin{figure}[h]
     \centering
     \includegraphics[width=\linewidth]{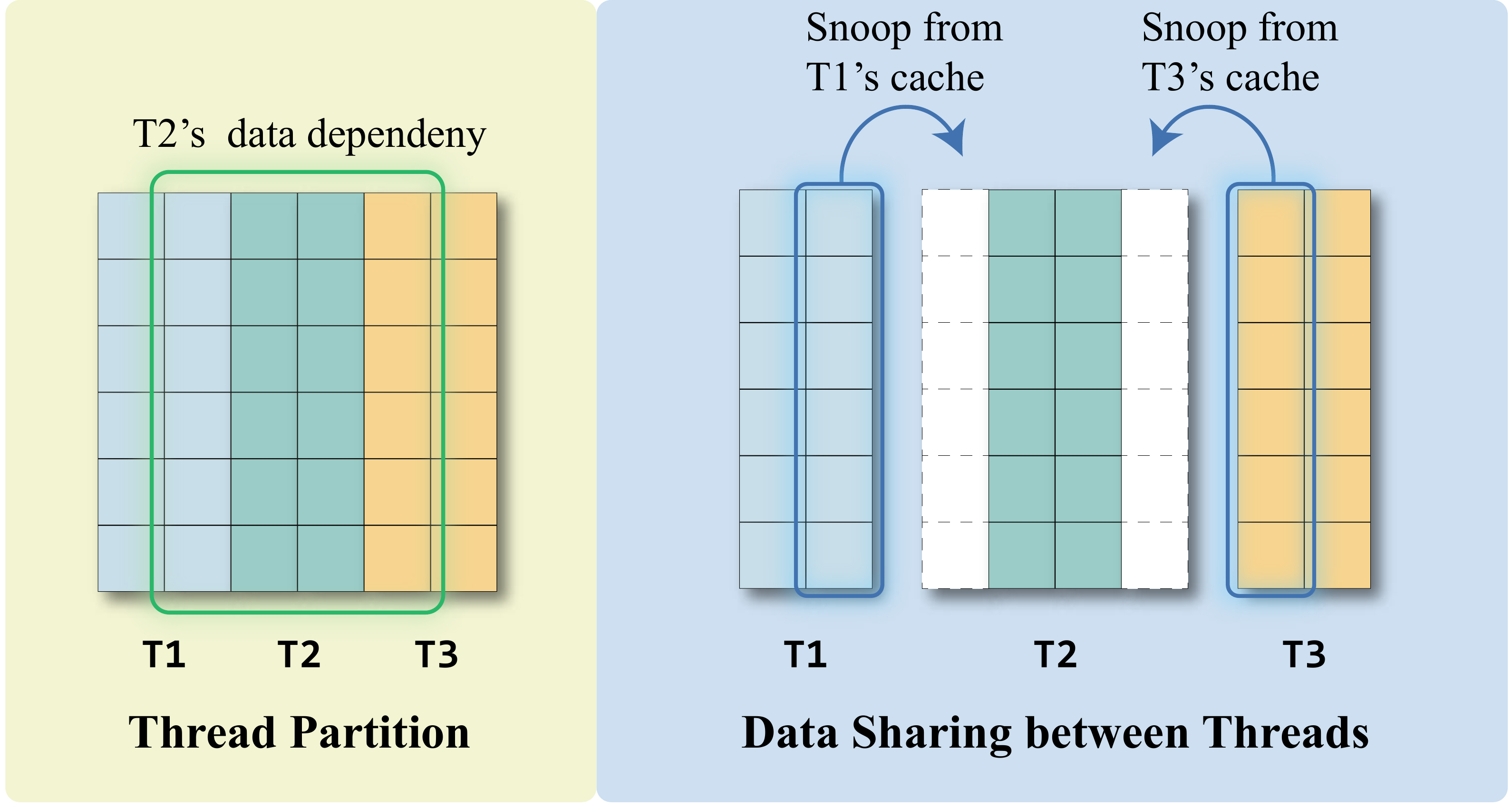}
     \caption{Cache-snoop based Data Sharing}
     \label{fig:l2-snoop}
\end{figure}

To address this limitation, we exploit the data cache coherence protocol. Whenever a core accesses an address missing in its private caches, it checks the NUMA's root directory. If the data resides in another core’s cache, it is retrieved efficiently through the intra-cluster interconnect. Leveraging this mechanism, we propose an \textit{cache-snoop based data sharing scheme} tailored for multicore CPU platforms. Specifically, as shown in Fig.~\ref{fig:l2-snoop}, we partition the grid into tiles that are narrower along the y-axis and spatially assign each core’s workload adjacently, enabling cores to access halo regions directly from neighboring cores rather than from main memory. Given that inter-core cache accesses are considerably faster than main memory accesses, the $Tile_Y$ component can be excluded from the reuse ratio analysis, resulting in a higher theoretical upper bound:
\begin{align*}
    &\max_{\text{Tile}_X, \text{Tile}_Y}\quad \frac{\text{Tile}_X}{(\text{Tile}_X + 2B_X)} \\
    &\text{s.t.} \quad (V_Z + 2B_Z)(\text{Tile}_X + 2B_X)(\text{Tile}_Y + 2B_Y) \leq SIZE_{LLC}
\end{align*}
Each core only needs to manage data reuse along the x- and z-axes, significantly improving data reuse in parallel stencil computation.

\subsection{Multi-Process Scope Optimizations}
To address NUMA effects on multi-core CPUs, we adopt an MPI-based approach for inter-NUMA parallelism and OpenMP for intra-NUMA parallelism. 

Even though MPI provides multi-thread support, all implementations of MPI runtime layer require a global lock to protect shared data structures, ensuring concurrency but not full parallelization. Consequently, assigning only one process per NUMA node does not fully saturate inter-NUMA bandwidth within a server node. Although MPI offers Remote Memory Access (RMA) interfaces in shared-memory environments, it cannot control the underlying memory properties on the mutlicore CPU platform, where both DDR and on-package memory are available.

To overcome these limitations, we employ the system-level SDMA engine integrated into the compute die of CPU to manage halo exchange across NUMA domains. Each SDMA engine on the die can perform asynchronous strided memory copies both within the die and between dies, offering more efficient data copying method.

\begin{figure}[h]
     \centering
     \includegraphics[width=\linewidth]{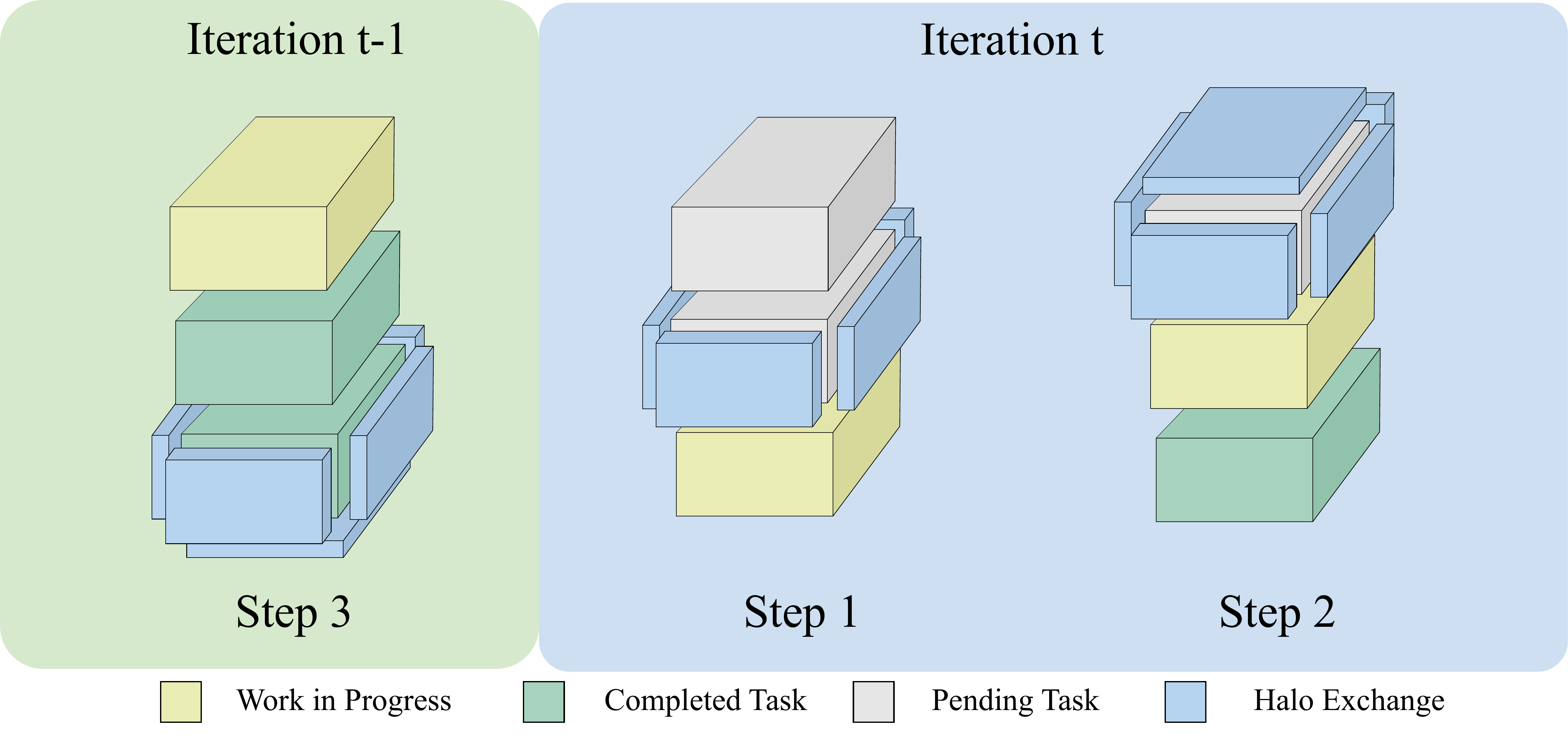}
     \caption{Pipeline Overlapping Scheme}
     \label{fig:pipeline}
\end{figure}

Another advantage of the SDMA engine is its non-intrusive nature: it does not occupy CPU cores or pollute caches. This allows efficient computation–communication overlap. Specifically, we apply a pipeline-overlap scheme\cite{Wang2020, Wan2023} by partitioning the grid into multiple layers along the $z$ axis, as shown in Fig.~\ref{fig:pipeline}. While computing the stencil on the current layer, we offload halo exchange tasks of next layer to the SDMA engine. Before moving to the next layer, we check for completion of the earlier SDMA tasks to resolve data dependencies. This approach further reduces communication overhead.

\subsection{Integrating MMStencil into HPC Applications}
In many HPC applications, different types of stencils are integrated into complex kernels, for example in earth modeling applications like RTM in complex media. To enable these applications to benefit from MMStencil, we provide basic stencil operators for computing small $(V_X, V_Y, V_Z)$ blocks. Complex kernels are decomposed to a series of small simple stencil operations on one $(V_X, V_Y, V_Z)$ block and then iterate through the entire domain. For kernels involving intermediate results, we leverage the CPU's private caches to allocate thread-private temporal buffers, which hold intermediate results for subsequent computations. As long as these buffers do not excessively occupy private cache space, data reuse of the input grid is preserved, avoiding unnecessary main memory traffic.

To illustrate, we present an example of accelerating RTM on TTI medium using MMStencil. This application requires computing six second-order partial derivatives:$
\frac{\partial^2 p}{\partial x^2}$, $\frac{\partial^2 p}{\partial y^2}$, $\frac{\partial^2 p}{\partial z^2}$, $\frac{\partial^2 p}{\partial x\partial y}$, $\frac{\partial^2 p}{\partial x\partial z}$, $\frac{\partial^2 p}{\partial y\partial z}$.

The first three derivatives are efficiently computed using 1D stencils. The last three derivatives are commutative, meaning the order of differentiation does not affect the result. This allows us to first compute the intermediate first-order derivatives $\frac{\partial p}{\partial z}$ and $\frac{\partial p}{\partial y}$ using 1D stencils. These intermediate results are then used in subsequent 1D stencil operations to obtain the second-order derivatives.

\begin{figure}[h]
     \centering
     \includegraphics[width=\linewidth]{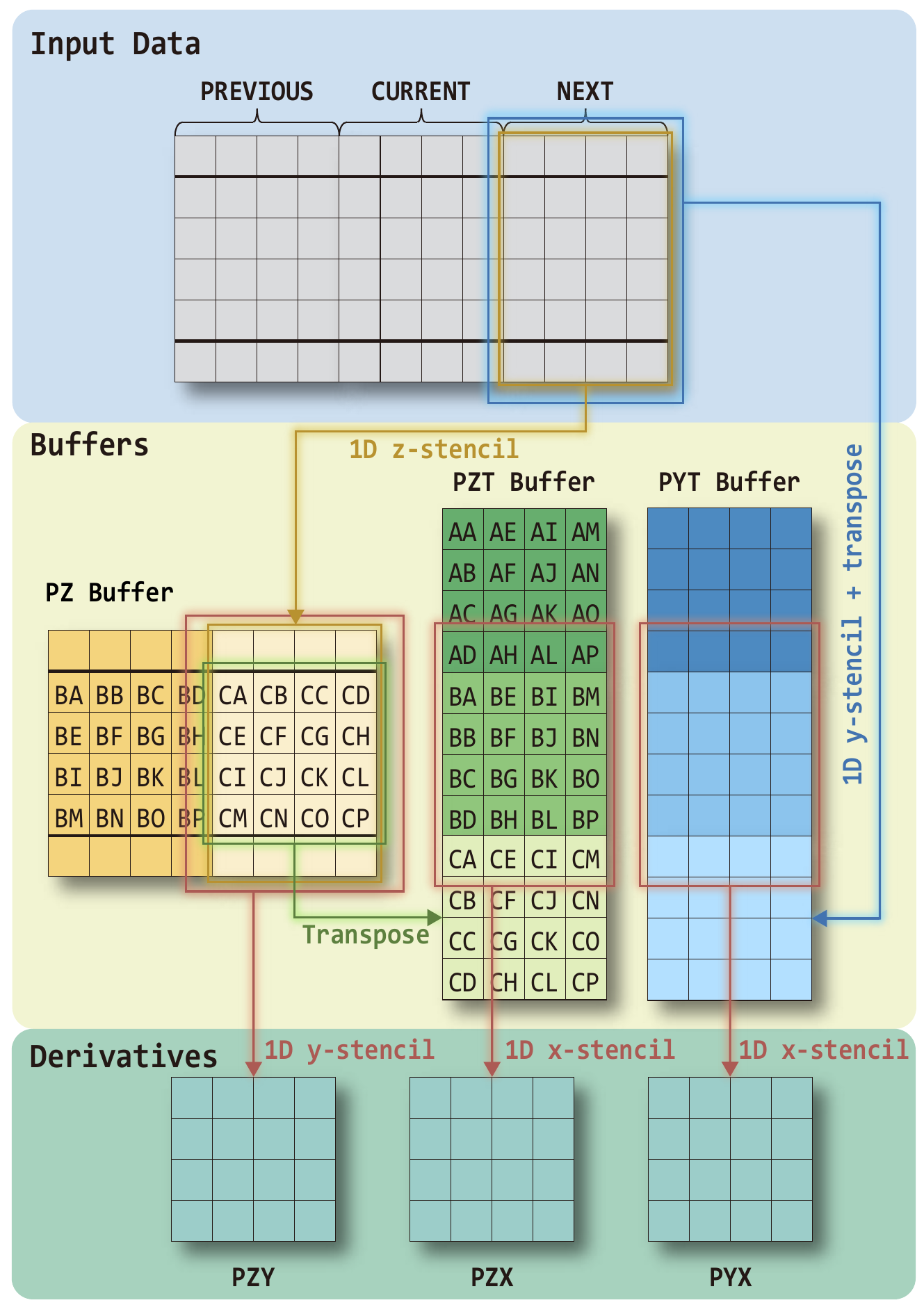}
     \caption{Computing partial derivatives}
     \label{fig:partial}
\end{figure}

As shown in Fig.~\ref{fig:partial}, to compute 
\( \frac{\partial^2 p}{\partial x \partial z}\) and 
\(\frac{\partial^2 p}{\partial y \partial z}\) for a block of size \((V_X, V_Y, V_Z)\), we proceed as follows:

\begin{enumerate}
  \item Perform the \(z\)–direction stencil on the “next” block of shape \((V_X, V_Y, V_Z)\) extended by a halo of size \((V_X, B_Y, V_Z)\) on both sides along the \(y\)–axis.
  \item To obtain \( \frac{\partial^2 p}{\partial x \partial z}\), transpose the intermediate \( \frac{\partial p}{\partial z}\) values from the the next block, then apply the \(x\)–direction stencil with data from the previous, and current block.
  \item To obtain \( \frac{\partial^2 p}{\partial y \partial z}\), apply the \(y\)–direction stencil to the intermediate \( \frac{\partial p}{\partial z}\) values from the current \((V_X, V_Y, V_Z)\) block together with the adjacent halo regions along the \(y\)–axis.
\end{enumerate}
The mixed derivative \( \frac{\partial^2 p}{\partial y \partial x}\) is computed analogously by composing a \(y\)–direction stencil followed by an \(x\)–direction stencil. Finally, scalar computations are accelerated using SIMD instructions to combine these derivatives with material properties and produce the final results.

\section{Experiments} \label{sec:experiments}

\begin{figure*}[h]
     \centering
     \includegraphics[width=\linewidth]{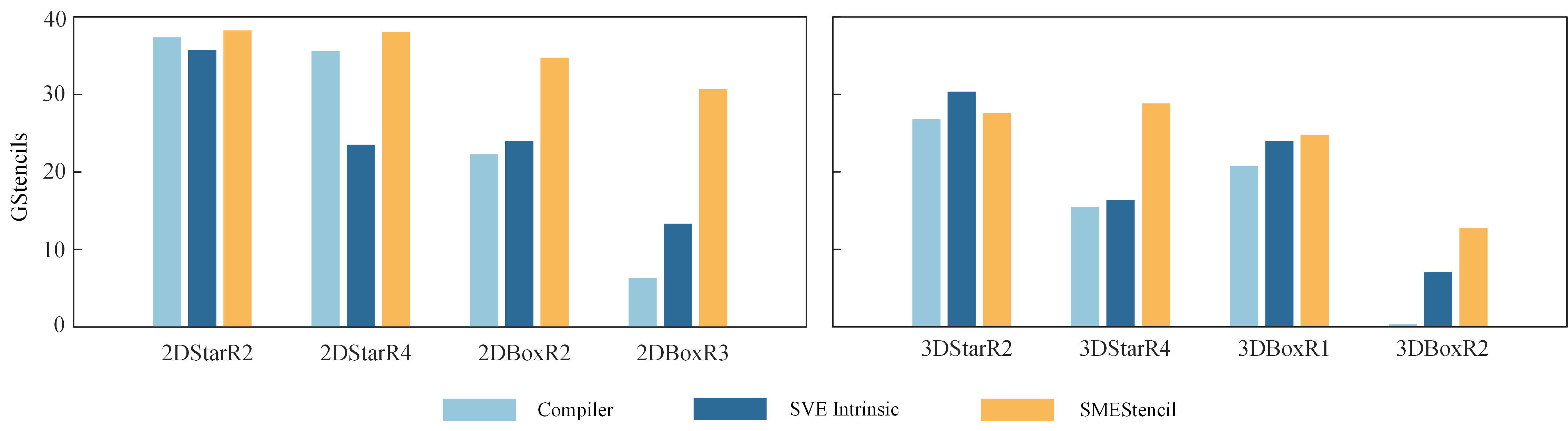}
     \caption{Performance Comparisons with Baselines}
     \label{fig:state-of-the-art}
\end{figure*}

\subsection{Experimental Setup} \label{sec:experimental_setup}

\paragraph{\textbf{Machine}}
Our experiments were conducted on the experimental platform introduced in \sect{arm9a}. Due to resource constraints, we utilized a single server node containing two processors, which collectively comprise 4 compute dies, 16 NUMA nodes, and 608 CPU cores. Unless performing multi-process scaling tests, all experiments are restricted to one on-package memory NUMA node.

\paragraph{\textbf{State-of-the-Art}}
Few stencil optimization approaches have been explored on our multicore CPU platform. Therefore, we implement a 2.5D blocking stencil using hand-written and manually loop-unrolled SIMD intrinsics combined with a $16\times4\times2$ brick layout to serve as our SIMD baseline.\footnote{We attempted to include BrickLib’s CPU implementation as one of our baselines; however, compatibility issues on our experimental platform prevented it from running correctly. Moreover, its binaries delivered lower performance compared to our hand-optimized implementation.}

\paragraph{\textbf{Benchmarks}}
We selected eight stencil benchmarks with varying shapes and dimensions, as detailed in Tab. \ref{tab:kernel_benchmark}. This set includes four star stencils (2DStarR2, 2DStarR4, 3DStarR2, 3DStarR4) and four box stencils (2DBoxR2, 2DBoxR3, 3DBoxR1, 3DBoxR2). Due to space limitations, we present results only for on-package memory and omit those for DDR. Additionally, Table~\ref{tab:kernel_benchmark} summarizes the key characteristics of these stencil kernels according to the CPU Roofline model \cite{Williams2009}, with respect to SIMD compute throughput and on-package memory bandwidth. We classify each kernel as follows:
\begin{itemize}
  \item \emph{Memory Bound}: performance is dominated by memory‐traffic overhead.
  \item \emph{Compute Bound}: CPU arithmetic throughput is the primary bottleneck.
  \item \emph{Both}: the kernel’s arithmetic intensity lies in the memory‐bound region but near the machine‐balance point, making it sensitive to both compute and memory limitations.
\end{itemize}

\begin{table}
  \caption{Stencil Kernel Benchmarks}
  \label{tab:kernel_benchmark}
  \begin{tabular}{cccc}
    \toprule
    Kernel & Points & Pattern & Tile Size \\
    \midrule
    2DStarR2 & 9   & Memory Bound   & (512, 512, 4) \\
    2DStarR4 & 17  & Memory Bound   & (512, 512, 4) \\
    2DBoxR2  & 25  & Memory Bound   & (512, 512, 4) \\
    2DBoxR3  & 49  & Both           & (512, 512, 4) \\
    3DStarR2 & 13  & Memory Bound   & (256, 16, 128) \\
    3DStarR4 & 25  & Memory Bound   & (256, 32, 64) \\
    3DBoxR1  & 27  & Memory Bound   & (256, 16, 128) \\
    3DBoxR2  & 125 & Computation Bound & (256, 16, 128) \\
    \bottomrule
  \end{tabular}
\end{table}

\subsection{Performance Breakdown}
In this subsection, we analyze the impact of the optimizations implemented in MMStencil using four representative examples: 3dStarR2, 3dStarR4, 3dBoxR1, and 3dBoxR2. These kernels were evaluated on a $512\times512\times512$ grid in single precision on both DDR memory and on-package memory.

\begin{figure}[h]
     \centering
     \includegraphics[width=\linewidth]{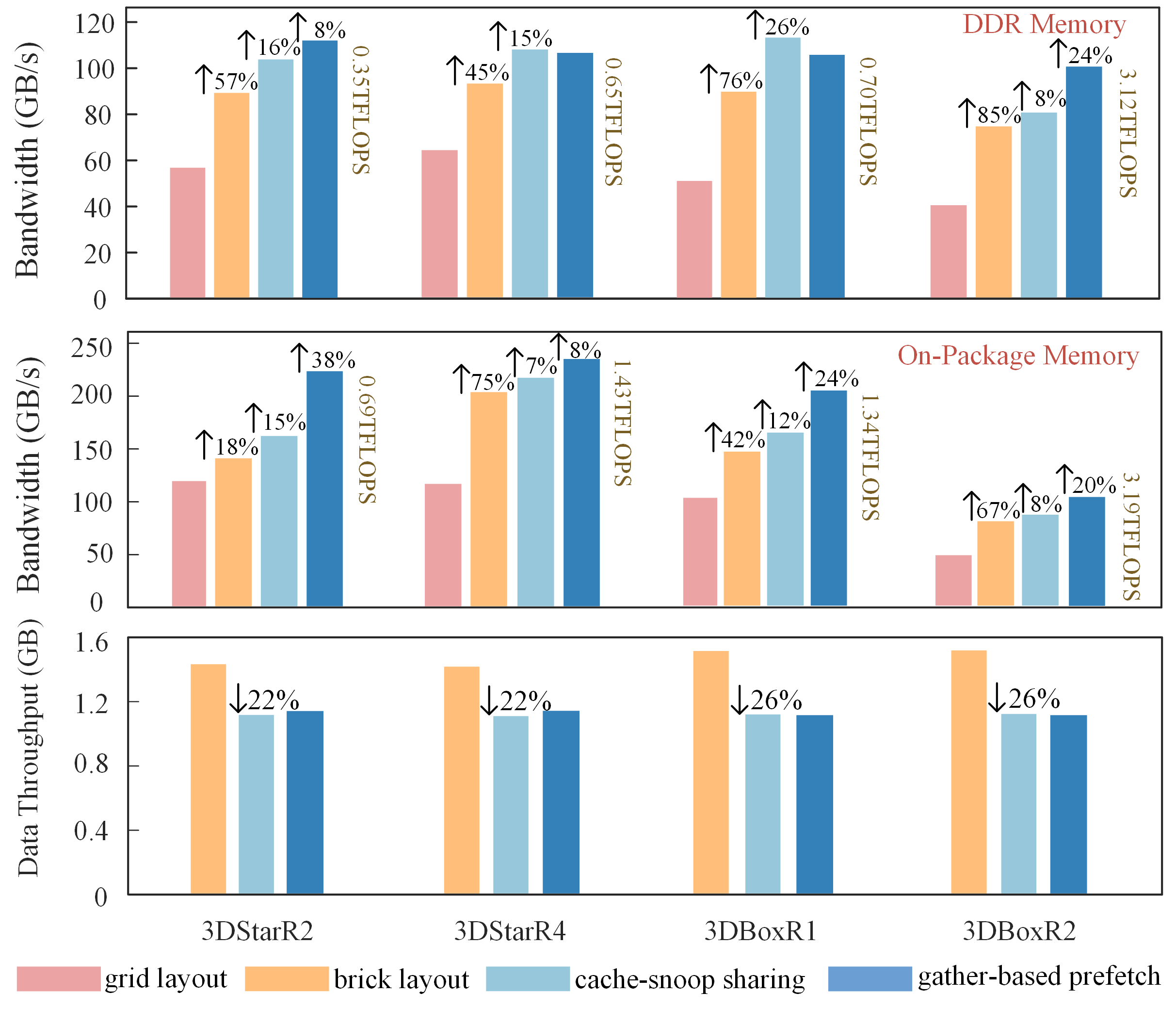}
     \caption{Performance Breakdown of MMStencil}
     \label{fig:performance-breakdown}
\end{figure}

As shown in Fig.~\ref{fig:performance-breakdown}, transforming the data layout from a regular grid to a brick layout yields the most significant performance gains on both DDR and on-package memory. Since most stencil kernels remain memory-bound, limiting the number of concurrent memory-access streams improves memory‐system efficiency and boosts overall performance.

The cache‐snooping scheme reduces main‐memory accesses by leveraging peer cores’ L2 caches to serve stencil halo regions. As shown in Fig.~\ref{fig:performance-breakdown}, this approach lowers global memory traffic by 22.12\%, 21.81\%, 26.17\%, and 26.17\% for the evaluated kernels, and delivers up to a 26\% performance improvement on DDR memory. However, on on‐package memory the gains are smaller: each core must still consult the root directory before retrieving data from another core’s cache, creating a bottleneck that limits the achievable speedup on on‐package memory.

The gather‐based prefetch scheme yields minimal performance benefits on DDR memory, except for the 3DBoxR2 kernel, where its high compute intensity allows better overlap of computation and memory access. This is because DDR bandwidth is relatively low compared to the Matrix unit’s compute throughput, and its 64-bit port can be easily saturated. In contrast, on on-package memory the gather‐based prefetch achieves additional performance gains of 38.09\%, 8.19\%, 24.26\%, and 19.74\% for the respective kernels. This optimization is particularly effective for stencils with short radii, as it trades extra memory traffic for more efficient, contiguous on-package accesses, thereby significantly improving bandwidth utilization.

\subsection{Comparison with State-of-the-Art Methods} \label{sec:smestencil_comp}

Fig.~\ref{fig:state-of-the-art} illustrates the performance comparison between MMStencil and other state-of-the-art approaches. 

For two‐dimensional stencils, the compiler‐optimized baseline handles star patterns efficiently, sustaining over 280 GB/s of effective bandwidth ($\approx 70\%$ of peak), leaving little headroom for further improvement—even a hand‐tuned SIMD version cannot outperform it. Nonetheless, MMStencil still delivers an 8\% speedup over the compiler variant. For the more complex box stencils, the compiler fails to generate high‐performance code: our SIMD version outpaces the compiler by 8\% and 112\% for radii 2 and 3, respectively. Building on the SIMD baseline, MMStencil leverages the matrix pipelines to achieve 1.44× and 2.31× speedups over the best CPU implementation for radius‐2 and radius‐3 box stencils, respectively, thereby validating our “Redundant Access Zeroing Box Stencil Optimization” scheme.

In the case of 3D stencils, the SIMD intrinsic version surprisingly delivers the best performance for the 3DStarR2 kernel. Although 3DStarR2 exhibits higher arithmetic intensity than 2DStarR2, MMStencil incurs additional overhead when switching from x- and y-axis stencils to the z-axis stencil. Moreover, since our multicore CPU core operates at a higher frequency in SIMD mode than in Matrix mode, the SIMD version can outperform MMStencil on simpler stencils. The improvement for the 3DBoxR1 kernel is modest compared to other CPU SIMD implementations, primarily because its short stencil radius prevents the Matrix unit from attaining a computational‐throughput advantage over the SIMD unit (see \sect{perf_model}).

Despite these modest gains for stencils with short radii, MMStencil achieves an average speedup of 80\% on high-order stencils compared to the best CPU implementations. Notably, for 3D high-order star stencils, performance even exceeds that of 3D star stencils with short radii. This can be attributed to the slower growth in the number of outer product operations relative to the number of stencil points, as well as the improved on-package memory access patterns of high-order stencils.

MMStencil consistently sustains high hardware utilization: for 2D star stencils, the effective bandwidth utilization exceeds 70\% across all radii; for 2D box stencils—despite their higher arithmetic workload—MMStencil maintains over 60\% utilization. In 3D star stencils, utilization reaches up to 57\%. Finally, for the compute‐bound 3DBoxR2 kernel , the theoretical peak throughput is 3.75TFLOPS(see \sect{perf_model}), and MMStencil attains 3.19TFLOPS—approximately 85\% of peak.  

\subsection{Discussion on MMStencil Performance}

Why does MMStencil demonstrate performance gains over the SIMD version while tensor‐core approaches failed? We attribute this improvement to the low instruction‐scheduling overhead of Matrix unit. The Matrix unit’s outer‐product model offers both low latency (only 4 cycles on our experimental platform) and high throughput(0.5 CPI in single-precision), whereas tensor cores incur latencies on the order of tens of cycles. Furthermore, our multicore CPU’s out‐of‐order execution resources allow the core to issue other instructions concurrently with outer‐products.

In comparison with SIMD version, by contrast, a CPU SIMD unit must execute two FMA instructions to achieve peak FLOPS, whereas Matrix unit requires only a single outer‐product instruction every two cycles to saturate its compute pipelines. This difference affords the CPU greater scheduling flexibility to interleave outer‐product operations with auxiliary tasks, such as vector loads, index calculations, and permutations, and to generate operands for multiple matrix accumulators in a pipelined fashion. As a result, MMStencil sustains high Matrix unit utilization and consistent performance, while SIMD‐based implementations suffer from instruction‐scheduling bottlenecks. Consequently, the SIMD version cannot attain its theoretical peak floating‐point throughput.

\subsection{Multi-process Experiments}
In this subsection, we evaluate the scaling behavior of MMStencil when the stencil computation is parallelized across different NUMAs. The default configuration employs the brick layout.

\begin{table}
  \caption{Halo Area Exchange Experiment}
  \label{tab:halo_exchange}
  \begin{tabular}{c|c|c|c}
    \toprule
    Direction & X & Y & Z \\
    \midrule
    Block Shape & (16, 512,512)   & (512, 4, 512)   & (512, 512, 4) \\
    MPI & 3.62GB/s  & 5.31GB/s   & 6.98GB/s \\
    SDMA  & 57.9GB/s  & 144.1GB/s   & 285.1GB/s \\
    Speedup  & 15.9$\times$  & 27.2$\times$        & 40.8$\times$ \\
    \bottomrule
  \end{tabular}
\end{table}

\begin{figure}[h]
     \centering
     \includegraphics[width=\linewidth]{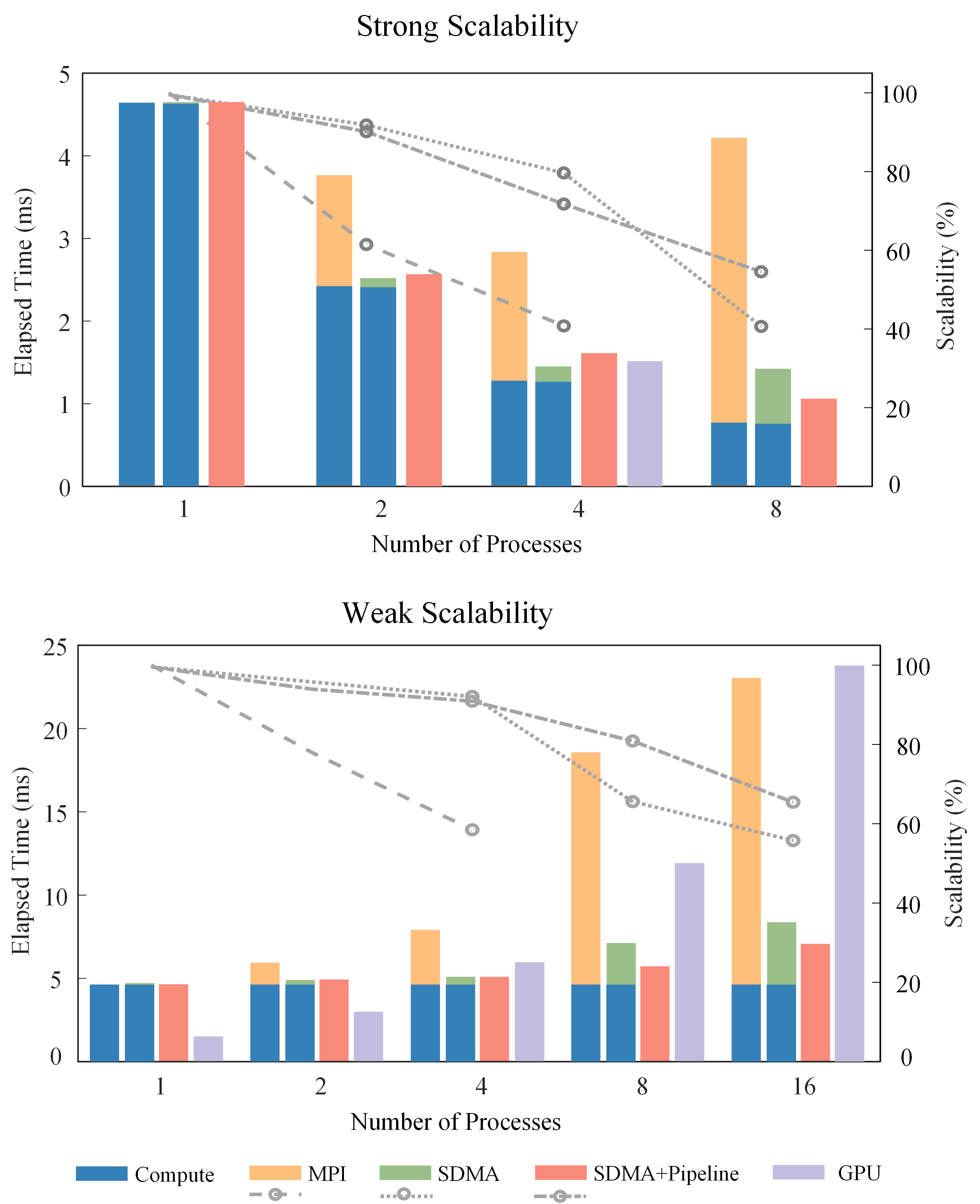}
     \caption{Scaling Experiments of MMStencil}
     \label{fig:scale}
\end{figure}

\subsubsection{SDMA Communication Bandwidth Test}
We first assess the data exchange bandwidth across NUMA nodes using both SDMA and MPI. The global grid size is set to $512\times512\times512$, and two processes are placed on the same DDR die to perform the halo exchange test. 

As shown in Tab. \ref{tab:halo_exchange}, the MPI implementation performs poorly, indicating that few processes cannot fully saturate inter-NUMA bandwidth. In contrast, SDMA achieves speedups of one to two orders of magnitude across all three directions. Despite these significant improvements, it is important to note that x-direction communication incurs a much higher cost compared to the y- and z-directions due to its more discontinuous memory access pattern and lower utilization of on-package bandwidth.

\subsubsection{Strong Scaling Test of Stencil}
For the strong scaling test, the limited on-package memory capacity per node constrains the problem size to $512\times512\times512$. 
The grid is partitioned along all three dimensions, with the Cartesian partitioning growing from $(1, 1, 1)$ to $(2, 2, 2)$. While avoiding partitioning in the x-direction could yield better scalability, we include it to ensure robustness even in the worst-case scenario.

As illustrated in Fig. \ref{fig:scale}, the MPI version is completely constrained by the overhead of halo exchange. The SDMA version exhibits much better scalability with up to 4 processes; however, when scaling to 8 processes, the unavoidable x-direction communication introduces overhead that prevents further performance gains. The pipeline version, which overlaps communication and computation, shows no significant improvement over SDMA at 4 processes, since the partitioned layers leave less workload for parallelization and the communication overhead is relatively minor. However, at 8 processes, where x-direction communication becomes more prominent, the pipeline overlapping scheme effectively yields further speedup.

To provide a clear comparison between MMStencil and SOTA CUDA implementation on an NVIDIA A100 GPGPU (80 GB), we include the elapsed time of BrickLib executing the 3DStarR4 stencil on the same domain size in single precision. When using half of the CPU resources (four NUMA domains), MMStencil matches the CUDA performance; when utilizing the entire CPU, despite the domain size being relatively small for full saturation, MMStencil still achieves a 1.5\(\times\) speedup over BrickLib.

\subsubsection{Weak Scaling Test of Stencil}
In the weak scaling test, each process handles a grid of size $512\times512\times512$. The number of processes is increased from $(1, 1, 1)$ to $(2, 2, 2)$ and finally to $(2, 2, 4)$.

As shown in Fig. \ref{fig:scale}, 
the SDMA version achieves near-ideal scalability with up to 4 processes; however, once x-direction communication is introduced at 8 processes, the overhead immediately degrades scalability. 
The pipeline version demonstrates better scalability when the number of processes exceeds 4, underscoring the importance of overlapping communication with computation when x-direction partitioning is unavoidable or when facing significant inter-processor communication. The suboptimal inter-processor communication becomes the major bottleneck with 16 processes.


Compared to BrickLib’s performance on the same domain size, MMStencil achieves a 1.2\(\times\) speedup when parallelized across four NUMA domains, and a 2.1\(\times\) speedup when utilizing all eight NUMA domains on a single CPU.

\begin{figure}[h]
     \centering
     \includegraphics[width=\linewidth]{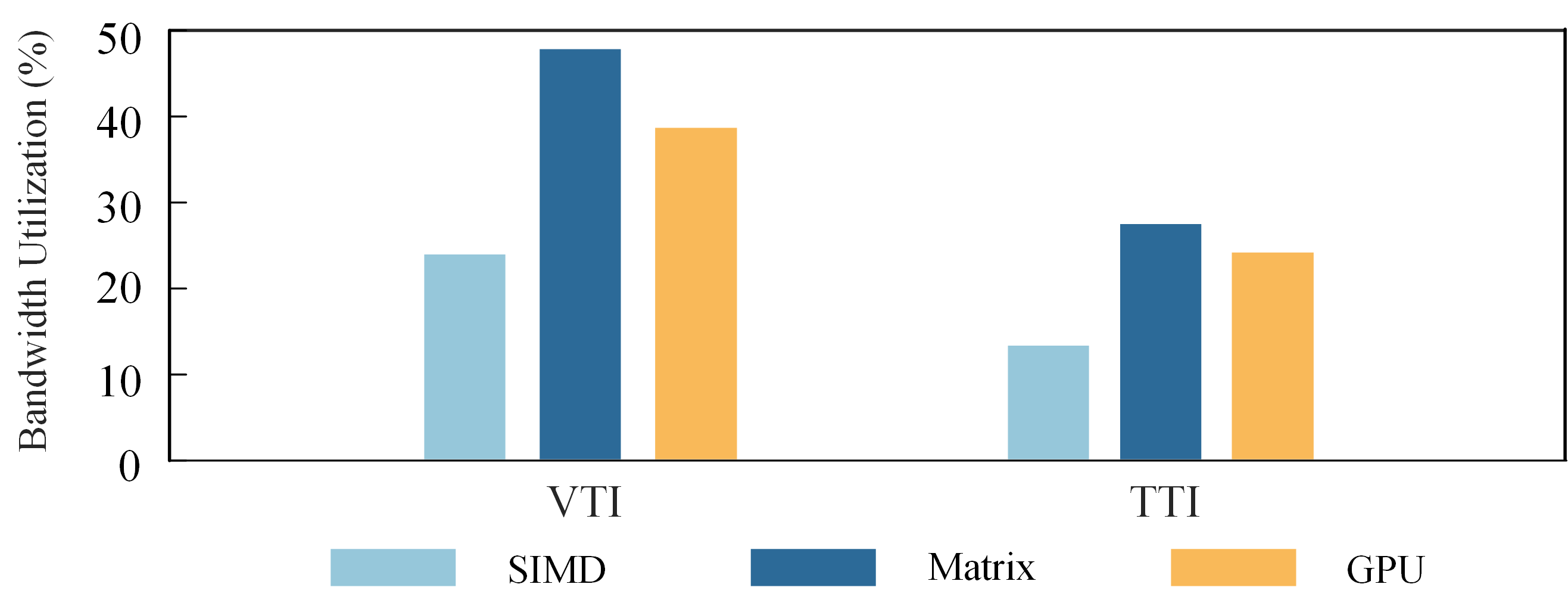}
     \caption{RTM Performance using MMStencil}
     \label{fig:rtm}
\end{figure}

\subsection{Performance in HPC Applications}

\begin{figure}[h]
     \centering
     \includegraphics[width=\linewidth]{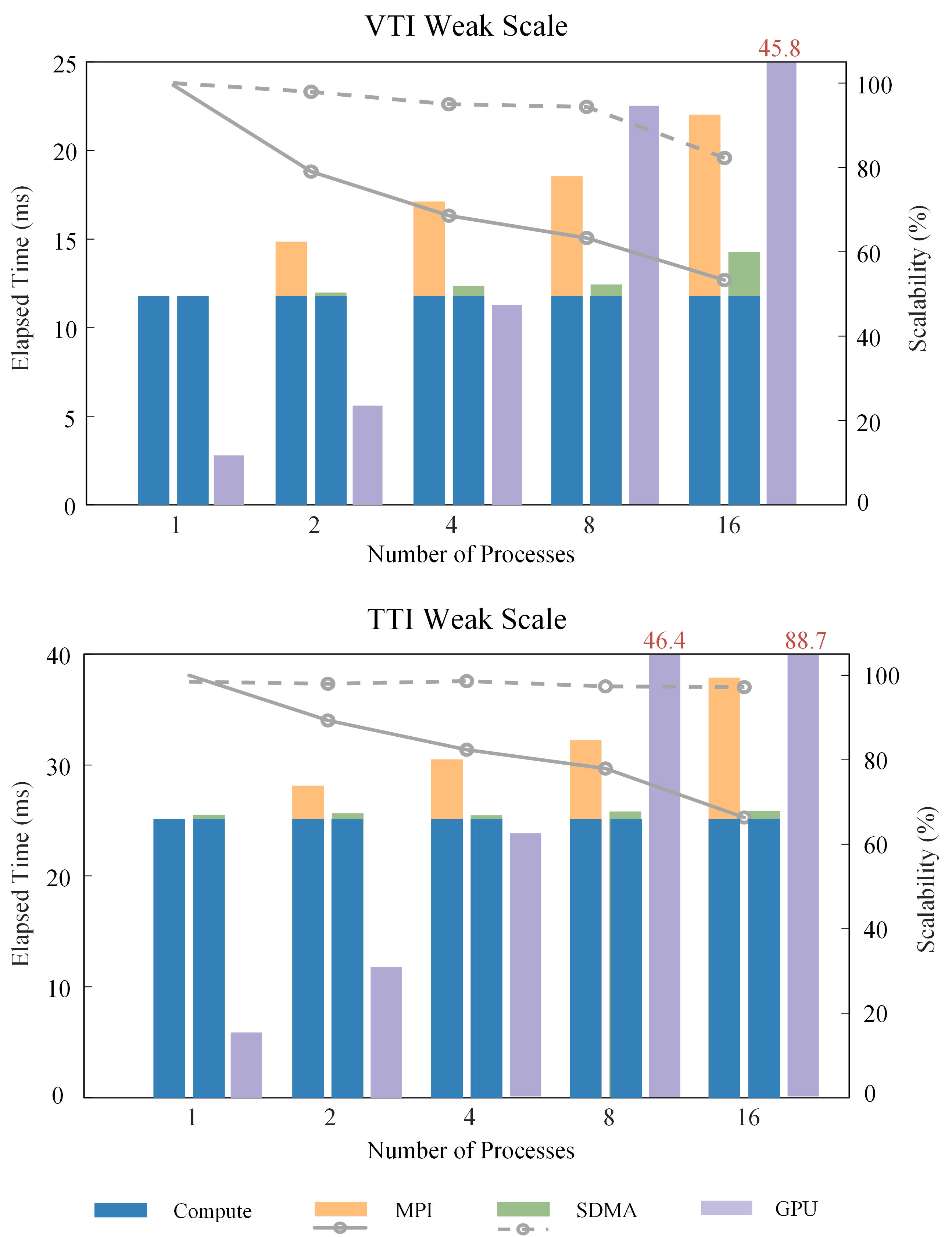}
     \caption{RTM Scaling Experiments}
     \label{fig:rtm_scale}
\end{figure}

In this subsection, we evaluate the performance gains achieved by integrating MMStencil into complex HPC applications. Leveraging our expertise in earth modeling, we target RTM in both VTI and TTI media. Our baselines consist of industrially optimized RTM implementations for both CPU and GPU, as provided by Bureau of Geophysical Prospecting INC. China National Petroleum Corporation. The CPU version runs on the experimental platform, while the GPU version is executed on an Nvidia A100 80GB on-package memory system. Due to on-package memory capacity limitations on the CPU, the grid size is set to $(512, 512, 256)$, whereas on the GPU it is $(512, 512, 512)$.

Fig. \ref{fig:rtm} shows the performance of RTM on VTI and TTI media using a single NUMA node. For the VTI case, the system achieves 47\% bandwidth utilization, a result consistent with that of the 3DStarR4 benchmark. Compared to the industrially optimized SIMD version, MMStencil achieves a 2.00$\times$ speedup. Moreover, when compared to the highly optimized GPU implementation, MMStencil demonstrates a 23.2\% improvement in bandwidth efficiency.

In the TTI case, where the temporal working set exceeds the L1 cache, the penalty for accessing intermediate data increases. Furthermore, decomposing the complex procedure into a sequence of one‐dimensional stencils amplifies the load/store overhead, reducing effective bandwidth utilization to 27.35\%—still on par with CUDA‐based implementations. A 2.06\(\times\) speedup over the SIMD version further validates the effectiveness of integrating MMStencil into complex HPC applications.

For scalability experiments of RTM, 
as shown in Fig. \ref{fig:rtm_scale}, replacing MPI with SDMA for inter-process communication significantly reduces the halo exchange overhead. When processes confined to a single processor, the communication overhead is negligible. Even when scaling up to 16 processes on two processors and incurring additional inter-processor communication overhead, this extra cost remains a small portion of the total execution time. We also compare against CUDA implementation executing the same workload. When scaling across four NUMA domains, MMStencil achieves runtimes comparable to the CUDA version; when utilizing both CPUs on a single server node, MMStencil delivers up to a 3.5\(\times\) speedup over the CUDA implementation.

\section{Conclusion}

We have presented MMStencil, a comprehensive optimization framework for three‐dimensional high‐order stencils on RISC multicore CPUs with the Matrix unit. Our approach spans microarchitectural tuning, memory‐layout transformations, multi‐threaded scheduling, and multi‐process NUMA‐aware communication.  

Through this study, we make four key observations. First, compiler‐generated code already achieves near‐peak performance on simple 2D and low‐order 3D stencils, suggesting that future efforts should prioritize more challenging patterns. Second, the high compute throughput of Matrix unit shifts the bottleneck back to memory, motivating further research into bandwidth‐efficient layouts and prefetch strategies especially on systems with on‐package memory. Third, decomposing complex 3D kernels into sequences of 1D stencils incurs nontrivial load/store overhead, indicating the need for more efficient decomposition schemes in real‐world applications. Finally, our results demonstrate that, with matrix accelerators and on‐package memory, CPUs can rival GPGPUs in stencil performance; extending these optimizations to additional HPC codes promises to unlock comparable gains on next‐generation CPU platforms.

\bibliographystyle{IEEEtran}
\bibliography{reference}

\begin{IEEEbiography}[{\includegraphics[width=1in,height=1.25in,clip,keepaspectratio]{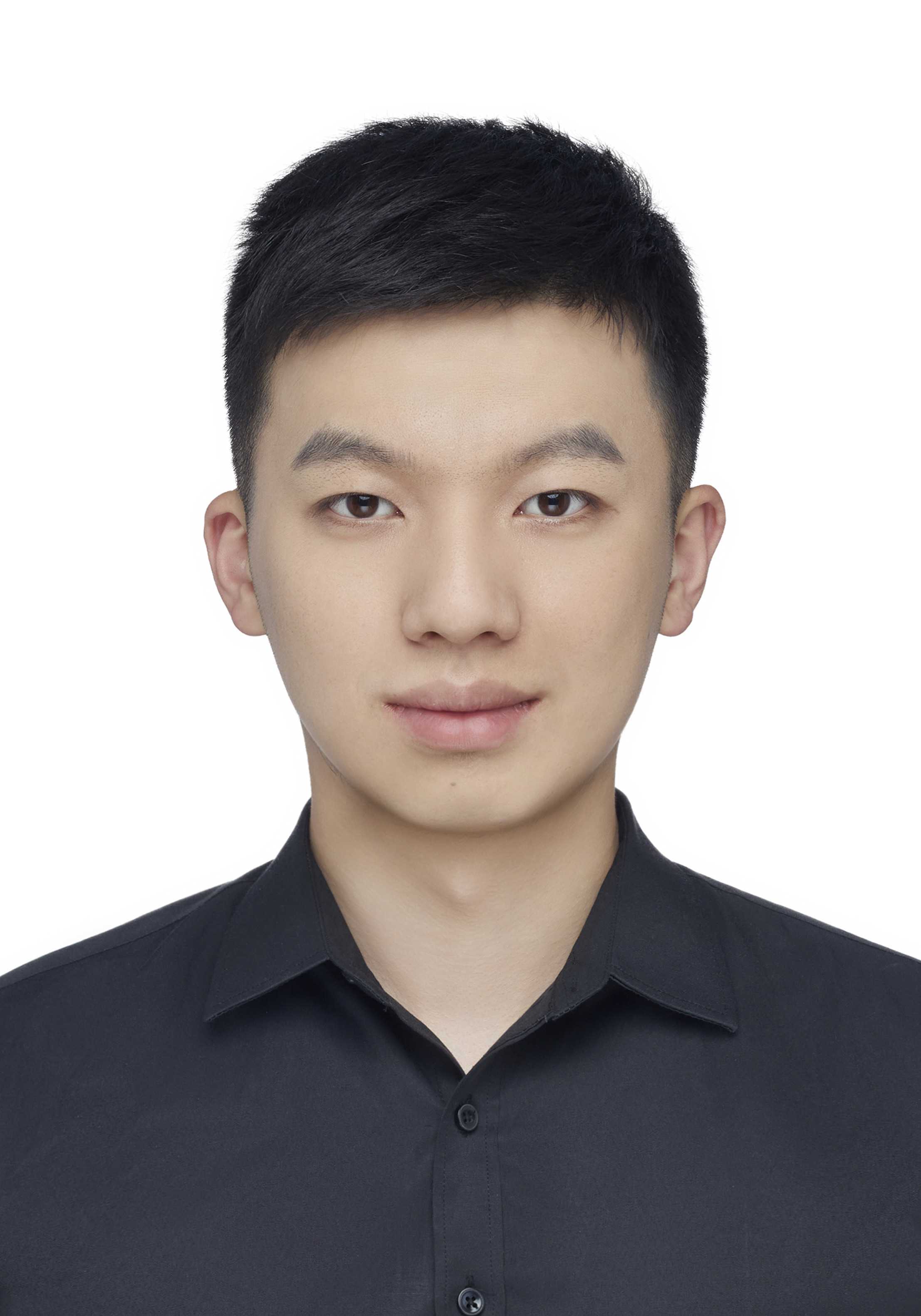}}]{
Yinuo Wang is a master candidate in the Department of Computer Science and Technology, at Tsinghua University. He focuses on the research of high-performance computing.}
\end{IEEEbiography}

\begin{IEEEbiography}[{\includegraphics[width=1in,height=1.25in,clip,keepaspectratio]{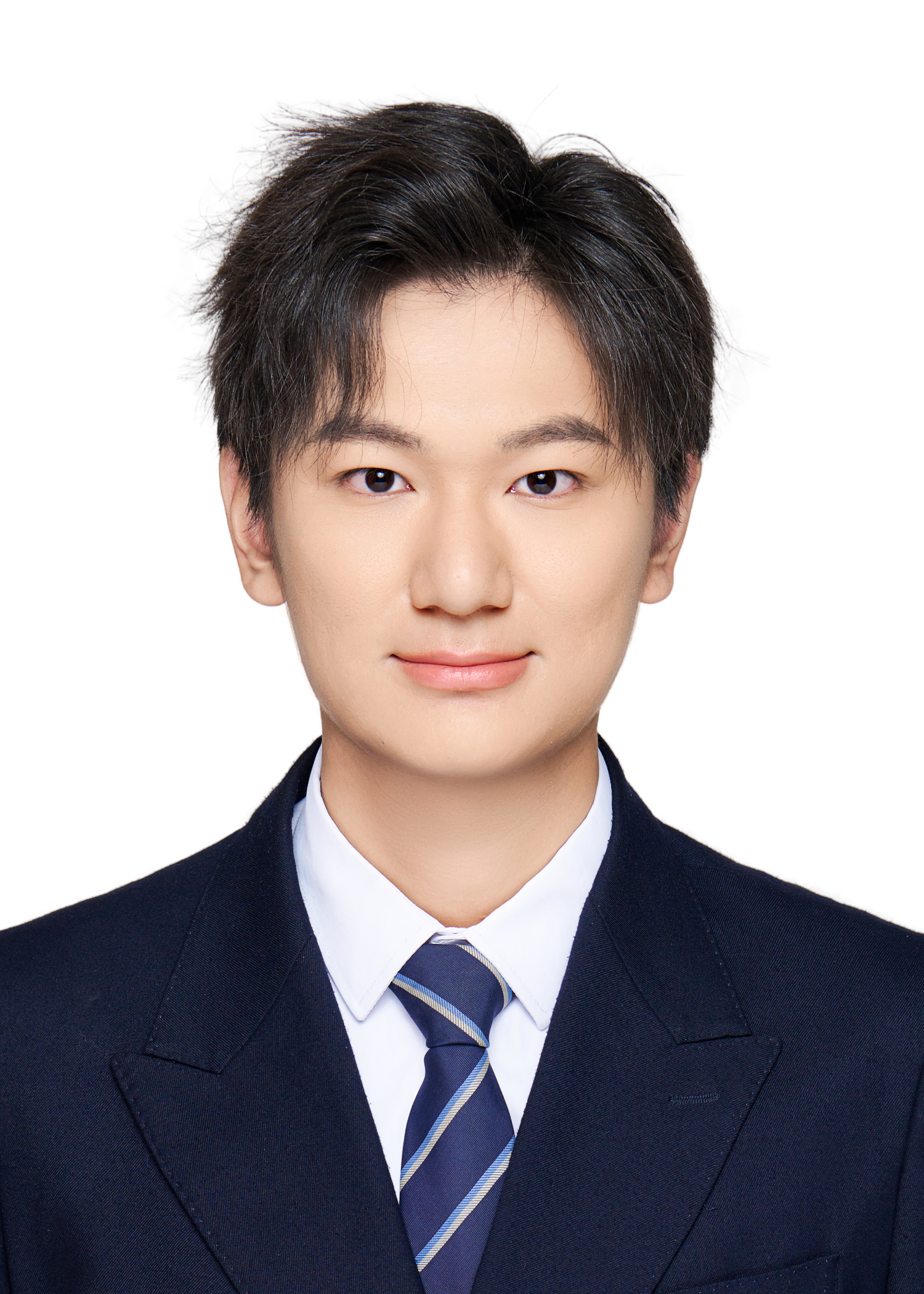}}]{
Tianqi Mao is a master candidate in the Department of Computer Science and Technology, at Tsinghua University. He focuses on the research of high-performance computing.}
\end{IEEEbiography}

\begin{IEEEbiography}[{\includegraphics[width=1in,height=1.25in,clip,keepaspectratio]{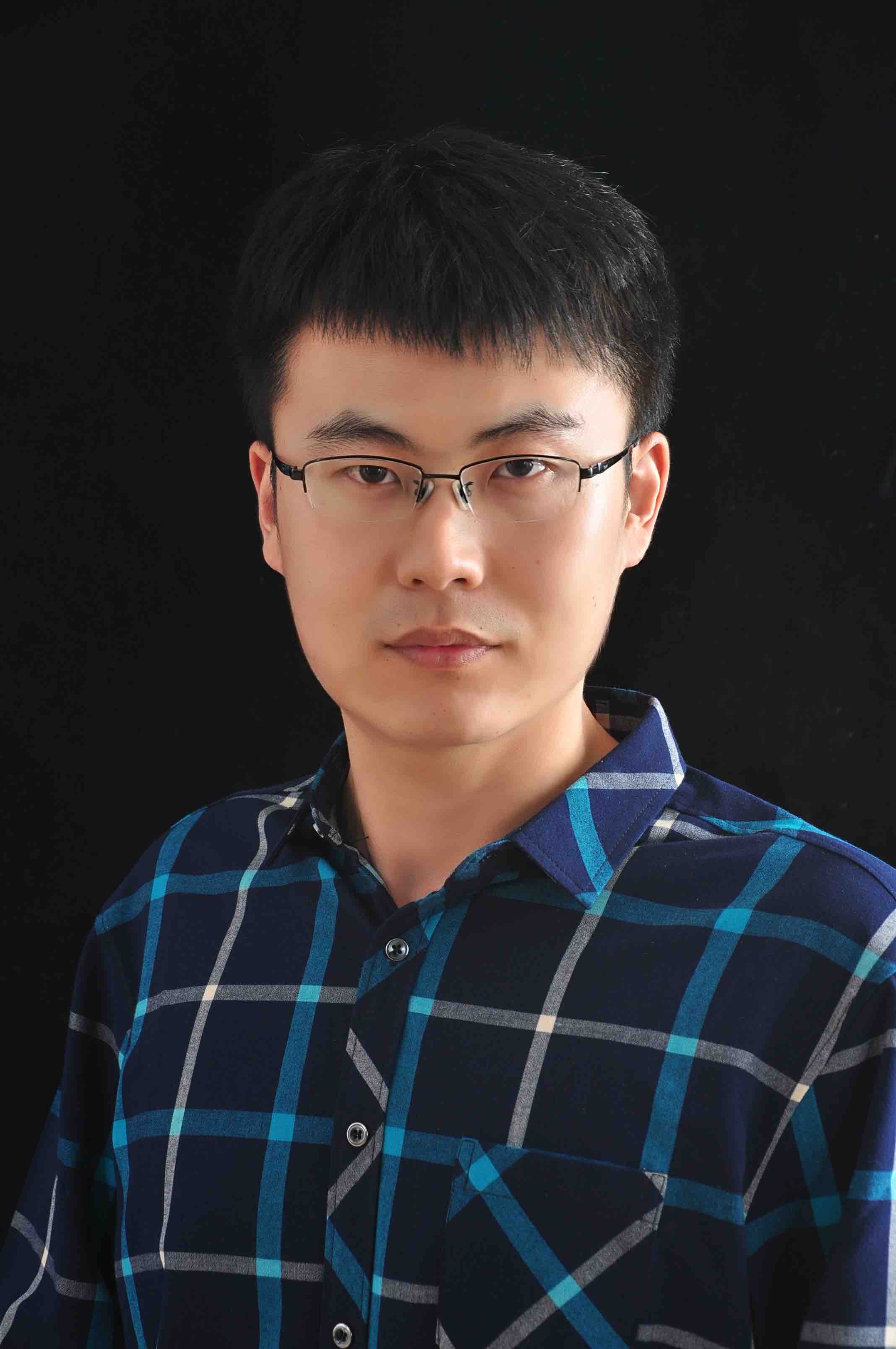}}]{
Lin Gan is an associate professor in Department of Computer Science and Technology, and Beijing National Research Center for Information Science And Technology, at Tsinghua University. He is also the assistant director of the National Supercomputing Center in Wuxi. His research interests include HPC solutions based on state-of-the-art systems. Gan has a PhD in computer science from Tsinghua University. He is the recipient of the 2016 ACM Gordon Bell Prize, the 2018 IEEE-CS TCHPC Early Career
Researchers Award for Excellence in HPC, the 2015 IEEE FPL Most Significant Paper Award in 25 Years, etc. He is a member of IEEE.}
\end{IEEEbiography}

\begin{IEEEbiography}[{\includegraphics[width=1in,height=1.25in,clip,keepaspectratio]{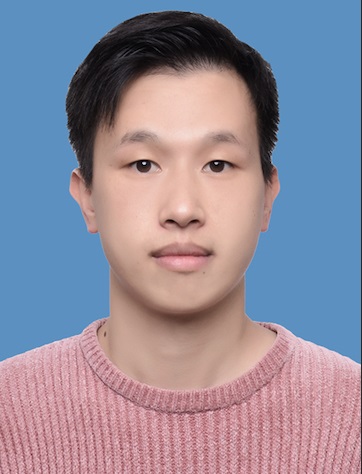}}]{
Wubing Wan is a Ph.D. candidate in the Department of Computer Science and Technology, at Tsinghua University. He focuses on the research of high-performance computing.}
\end{IEEEbiography}

\begin{IEEEbiography}[{\includegraphics[width=1in,height=1.25in,clip,keepaspectratio]{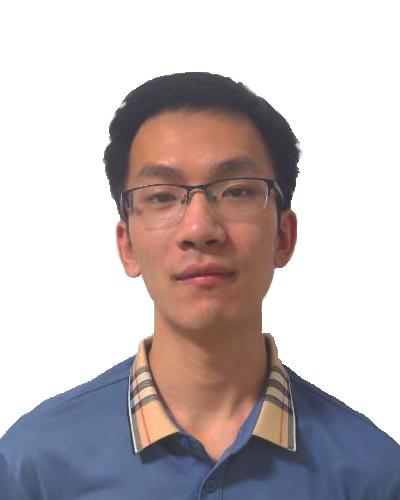}}]{
Zeyu Song received the Bachelor's degree in Computer Science from Tsinghua University. He is pursuing a Ph.D. in the Department of Computer Science and Technology at the same institution. His primary focus of research centers around HPC application acceleration on heterogeneous supercomputer platforms.}
\end{IEEEbiography}

\begin{IEEEbiography}[{\includegraphics[width=1in,height=1.25in,clip,keepaspectratio]{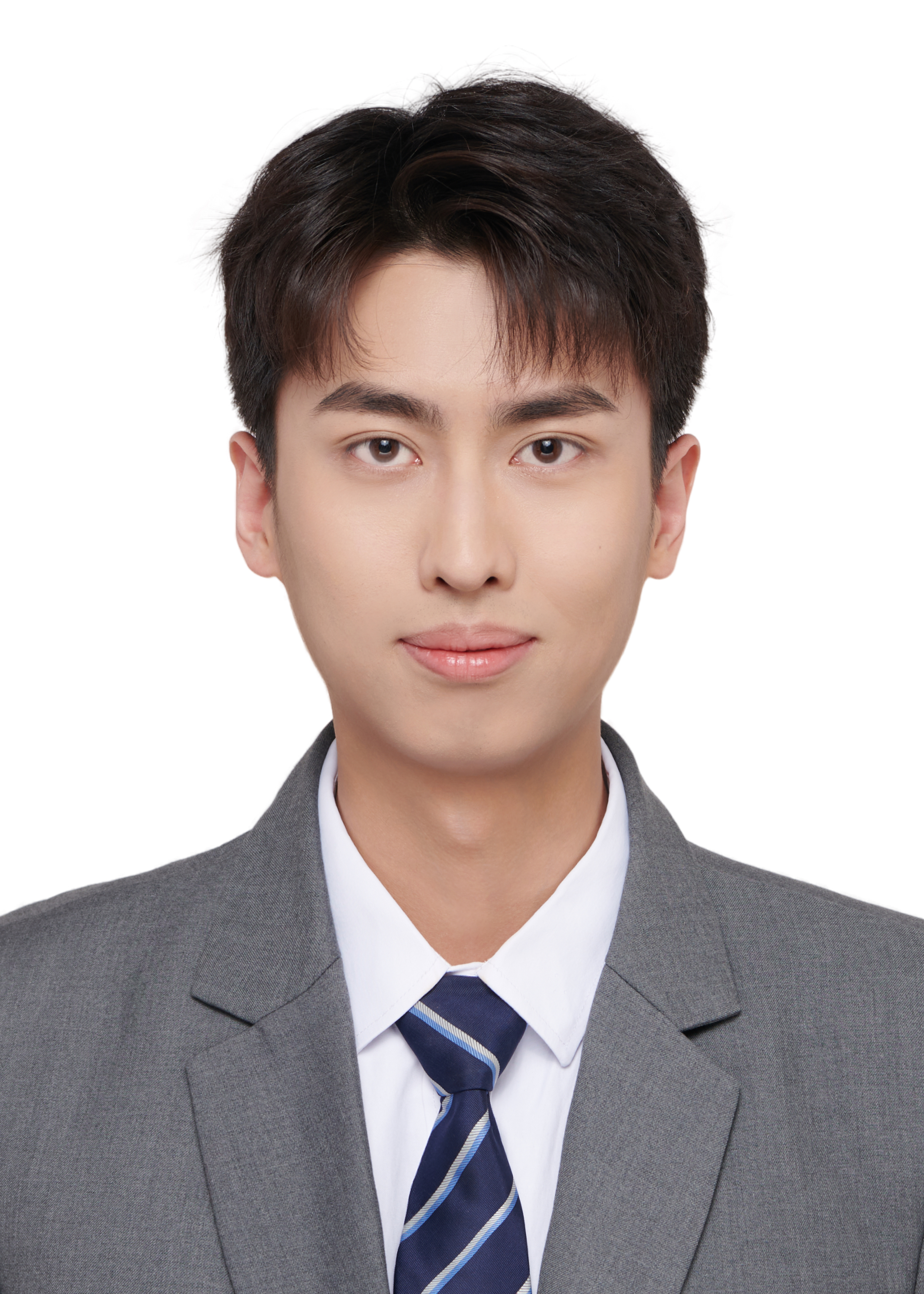}}]{
Jiayu Fu is a Phd. candidate in the Department of Computer Science and Technology, at Tsinghua University. He focuses on the research of high-performance computing.}
\end{IEEEbiography}

\begin{IEEEbiography}[{\includegraphics[width=1in,height=1.25in,clip,keepaspectratio]{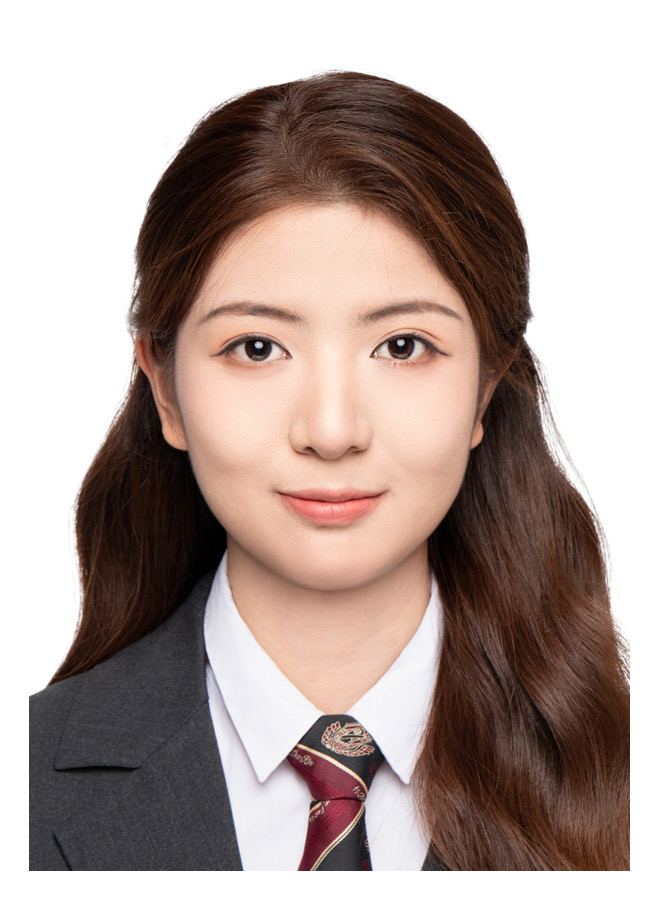}}]{
Lanke He obtained her master degree in the Department of Electrical Engineering, at Tsinghua University in 2025. Her research focuses is on power system modeling.}
\end{IEEEbiography}

\begin{IEEEbiography}[{\includegraphics[width=1in,height=1.25in,clip,keepaspectratio]{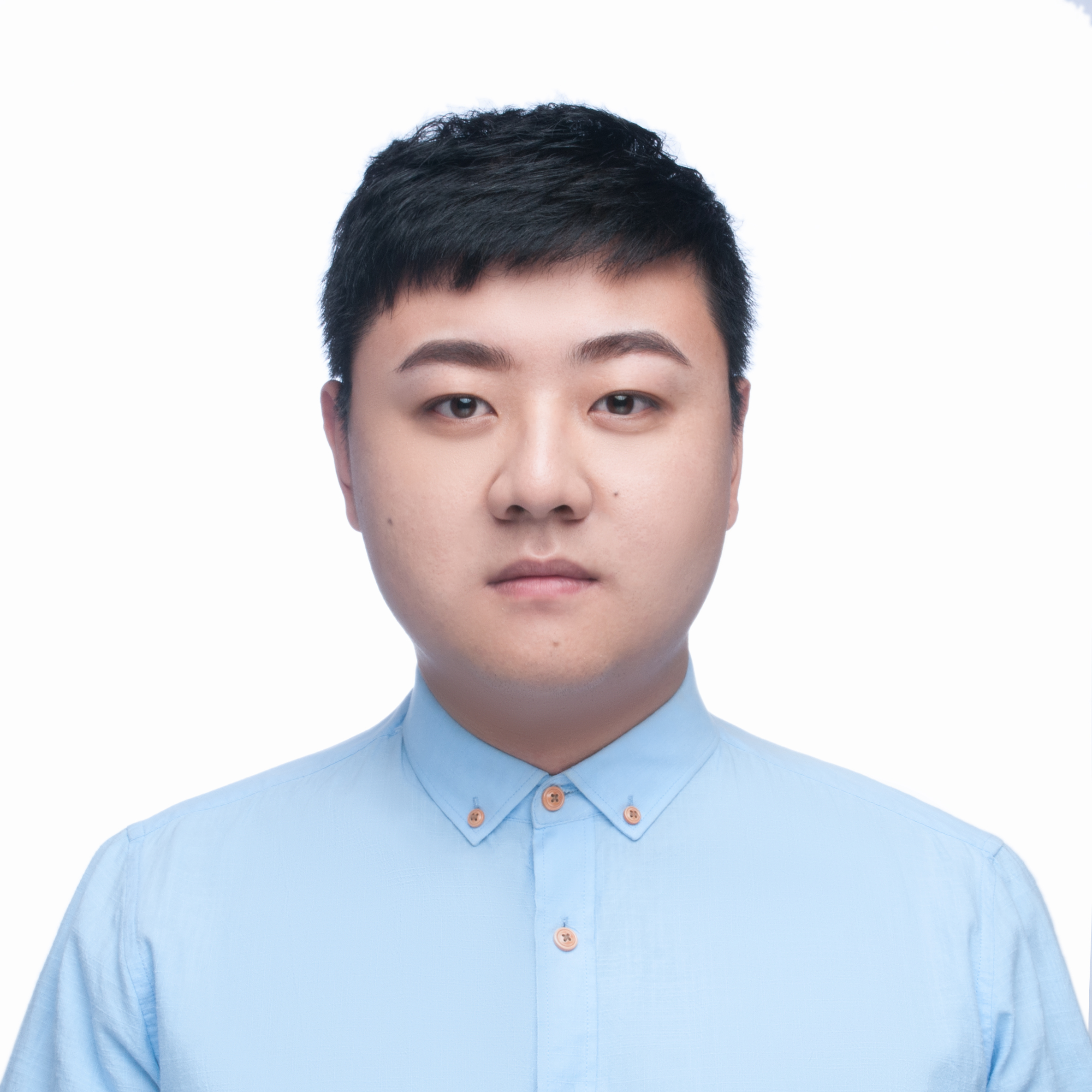}}]{Wenqiang Wang obtained his Ph.D. degree in Mechanics from Harbin Institute of Technology, China, in 2023. During his Ph.D. studies, he was a joint Ph.D. student at Southern University of Science and Technology, China. His research interests include seismic numerical algorithms, heterogeneous parallel computing, and 3D complex earthquake modeling.
}
\end{IEEEbiography}

\begin{IEEEbiography}
[{\includegraphics[width=1in,height=1.25in,clip,keepaspectratio]{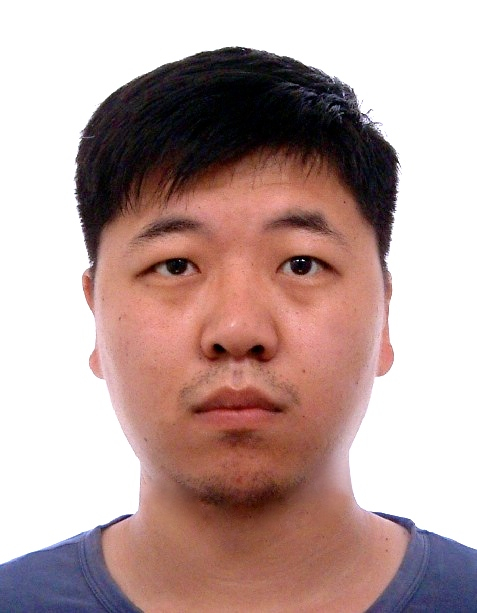}}]{Zekun Yin}
is an associate research professor at the School of Software, Shandong University. He received his bachelor’s and Ph.D. degrees from Shandong University in 2014 and 2020, respectively. His major research interests are high-performance computing and its application in life science and geoscience.
He won the ACM Gordon Bell Prize in 2017 for his research work on the Tangshan earthquake simulation on the Sunway TaihuLight supercomputer.
\end{IEEEbiography}

\begin{IEEEbiography}[{\includegraphics[width=1in,height=1.25in,clip,keepaspectratio]{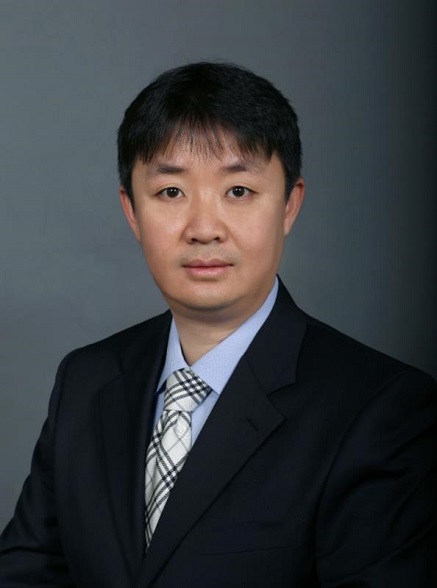}}]{
Dr. Wei Xue is a professor at the Department of Computer Science and Technology in Tsinghua University, China. He is the director of the High-Performance Computing Institute at Tsinghua and a joint faculty in the Department of Earth System Science at Tsinghua. His research interests include scientific computing, performance evaluation, and optimization, and uncertainty quantification. As one of the team leaders, he received the 2016 and 2017 Gordon Bell Prizes and finalist of the 2018 Gordon Bell Prize. He is a senior member of CCF and a member of IEEE and ACM.}
\end{IEEEbiography}

\begin{IEEEbiography}[{\includegraphics[width=1in,height=1.25in,clip,keepaspectratio]{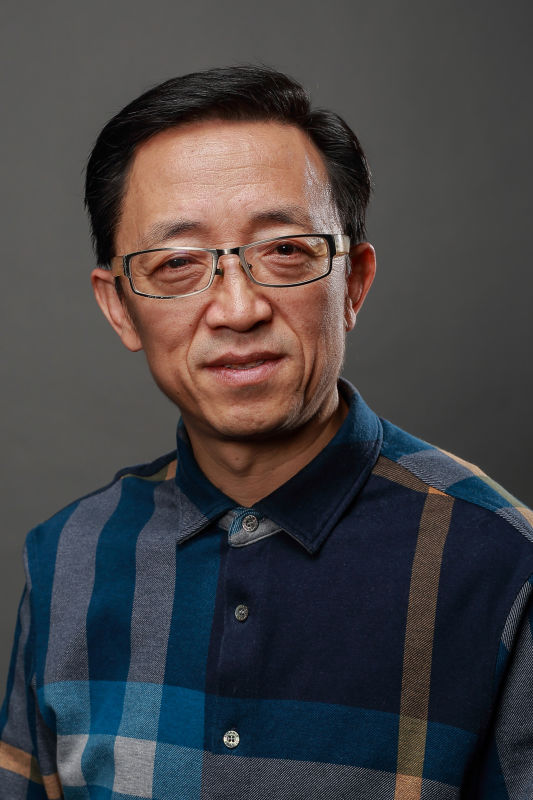}}]{
Guangwen Yang is a professor at the Department of Computer Science and Technology, and the Beijing National Research Center for Information Science and Technology, at Tsinghua University, China. He is also the director of the National Supercomputing Center in Wuxi, China. His research interests include parallel algorithms, cloud computing, machine learning, and the earth system model. Yang has a Ph.D. in computer science from the Harbin Institute of Technology. He has been awarded the ACM Gordon Bell Prize (2016, 2017). He is a member of IEEE.}
\end{IEEEbiography}

\vfill

\end{document}